\documentclass{elsarticle}
\usepackage{epsfig}
\def\beq{\begin{equation}}
\def\beqn{\begin{eqnarray}}
\def\eeq{\end{equation}}
\def\eeqn{\end{eqnarray}}

\def\bl{\bar l}
\def\bs{\bar s}
\def\bt{\bar t}
\def\bu{\bar u}
\def\bx{\bar x}
\def\dsb{d\bar\sigma}

\newcommand\sss{\scriptscriptstyle}
\newcommand\as{\alpha_{\sss S}}
\newcommand\aem{\alpha_{\sss em}}

\newcommand\Qb{\bar{Q}}

\newcommand\xMCB{\Big|_{\sss {\rm MC}}}

\newcommand\stepf{\Theta}
\newcommand\Hone{(e)}
\newcommand\Htwo{(H)}

\renewcommand\arraystretch{1.1}
\begin{document}
\begin{frontmatter}
%%%%CERN-PH-TH/2011-128
\title{Charm and bottom photoproduction at HERA with MC@NLO}
\author{T. Toll}
\ead{ttoll@bnl.gov}
\address{Physics Department, Brookhaven National Laboratory, Upton, NY 11973, USA}

\author{S. Frixione}
\ead{Stefano.Frixione@cern.ch}
\address{
  PH Department, TH Unit, CERN, CH-1211 Geneva 23, Switzerland\\
  ITPP, EPFL, CH-1015 Lausanne, Switzerland\\ }

%%%%ABSTRACT
\begin{abstract}
We apply the MC@NLO formalism to the production of heavy-quark
pairs in pointlike photon-hadron collisions. By combining this
result with its analogue relevant to hadron-hadron collisions, 
we obtain NLO predictions matched to parton showers for the 
photoproduction of $Q\bar{Q}$ pairs. We compare MC@NLO results
to the measurements of $c$- and $b$-flavoured hadron observables
performed by the H1 and ZEUS collaborations at HERA
\end{abstract}

\begin{keyword}
MC@NLO \sep heavy quarks \sep photoproduction
\end{keyword}

\end{frontmatter}

\section{Introduction}
The production of pairs of quarks with mass much larger than
the typical hadronic scale, thus called heavy, has distinctive features 
that renders it an interesting case study. From the theoretical viewpoint, 
the quark mass cuts the collinear singularities off, and open-quark
cross sections, in which no fragmentation functions or jet-finding
algorithms are used, are well defined in perturbation theory.
On the experimental side, heavy-quark production can be exploited
thanks to its peculiar signatures, and generally large rates.
The definition of what is heavy is to a certain extent ambiguous,
and may depend on the kinematic region one is interested into probing
(since the behaviour of a particle with mass $m$ can be fairly
different according to whether $m\gg p_T$ or $m\ll p_T$).
While there is no doubt that the top is heavy, the heaviness 
of the bottom quark may be debatable, and that of the charm 
quark certainly is. It remains true, however, that open charm
and bottom cross sections can be computed at fixed order in QCD;
the interesting question is therefore how predictions compare to data.

This question has indeed received a lot of attention in the past
twenty years, in part because of the important role played by heavy 
quarks in many discovery channels of BSM physics. Apart from the
mainstream activity at hadron machines, the $ep$ collider HERA has 
also been collecting charm and bottom data for over a decade now, in 
both the DIS and the photoproduction regimes (see e.g.~ref~\cite{Jung:2009eq}
for a review). With its relatively small c.m.~energy of about 300~GeV,
heavy-quark physics at HERA very rarely involves large transverse
momenta, and therefore offers a good testing ground for purely-perturbative
QCD predictions. It is now customary to compare data to results
accurate to NLO (${\cal O}(\aem\as^2)$)~\cite{Jung:2009eq};
fully exclusive computations, in both DIS~\cite{Harris:1997zq}
and photoproduction~\cite{Frixione:1993dg}, have been available
for some time now.

Comparisons between data and parton-level, fixed-order predictions
are not entirely satisfactory. They may work well for fully inclusive
observables dominated by hard scales, but unfortunately experimental
measurements typically involve hadron-level, smallish-scale quantities which 
cannot be well described by such a simple theoretical framework, and
corrections (which introduce biases) need be applied to theoretical
results, data, or both, in this way blurring the picture.
Parton Shower Monte Carlos (PSMCs) offer a viable alternative, with
their fully realistic final states, but lack the accuracy of
higher-order perturbative computations. The solution to the problem
of matching the two approaches has been extensively studied in the 
past ten years, and is now fairly well established (see 
e.g.~ref.~\cite{Salam:2010zt} for a pedagogical introduction).

The purpose of this paper is that of applying the MC@NLO matching
formalism~\cite{Frixione:2002ik} to the case of heavy-quark
pair photoproduction, and of comparing the (hadron-level) results
obtained in this way with selected bottom and charm measurements
performed by the H1 and ZEUS collaborations at HERA.
This paper is organized as follows: in sect.~\ref{sec:gen} we give
the briefest introduction to the photoproduction jargon; in 
sect.~\ref{sec:sub} we summarize a few technical information on MC@NLO,
specific to photon-initiated processes. Section~\ref{sec:meas}
presents the comparisons between theoretical predictions and
data; finally, in sect.~\ref{sec:concl} we report our conclusions.

\section{Photoproduction: generalities\label{sec:gen}}
The high-energy photons that initiate photoproduction processes are
typically obtained by quasi-collinear bremsstrahlung off an
electron beam. We write the factorization theorem relevant 
to photoproduction as follows:
\beqn
d\sigma_{eH}&=&\sum_b\int dx_1 dx_2
f_{\gamma}^{\Hone}(x_1)f_b^{\Htwo}(x_2)
d\sigma_{\gamma b}(x_1,x_2)
\nonumber \\*&+&
\sum_{ab}\int dx_1 dx_2
f_{a}^{\Hone}(x_1)f_b^{\Htwo}(x_2)
d\sigma_{ab}(x_1,x_2).
\label{factth}
\eeqn
We have denoted by $f_{\gamma}^{\Hone}$ the Weizs\"acker-Williams 
function~\cite{von Weizsacker:1934sx,Williams:1934ad}, which gives a 
good approximation of the energy spectrum of photons radiated by the 
electrons (for the computations of this paper, we have used the 
form presented in ref.~\cite{Frixione:1993yw}). The functions 
$f_{a}^{\Hone}$, the electron PDFs, are the analogues of the hadron PDFs, 
and are defined as follows:
\beq
f_{a}^{\Hone}(x)=\int dy dz \delta(x-yz)
f_{\gamma}^{\Hone}(y)f_a^{(\gamma)}(z)\,,
\eeq
with $f_a^{(\gamma)}$ the relevant photon PDFs. The two terms on the r.h.s.
of eq.~(\ref{factth}) are usually called the pointlike and the hadronic
photon components respectively. As the name suggests, the hadronic 
component is computed in exactly the same way as any ordinary
hadron-hadron cross section. MC@NLO is no exception to this rule,
and we have therefore used the calculation of ref.~\cite{Frixione:2003ei},
with trivial modifications in the corresponding computer code that allow
the use of electron PDFs. In this work, we have performed the computation
of the pointlike contribution to $Q\Qb$ production in MC@NLO. The
underlying partonic processes are:
\beqn
\gamma + g&\longrightarrow& Q+\Qb\,,
\label{procBpg}
\\
\gamma + g&\longrightarrow& Q+\Qb+g\,,
\label{procRpg}
\\
\gamma + q&\longrightarrow& Q+\Qb+q\,,
\label{procRpq}
\eeqn
where eq.~(\ref{procBpg}) receives contributions from the Born 
(${\cal O}(\aem\as)$) and one-loop (${\cal O}(\aem\as^2)$) matrix elements, 
and eqs.~(\ref{procRpg})--(\ref{procRpq}) receive contributions from the 
real-emission matrix elements (${\cal O}(\aem\as^2)$), and from the 
corresponding counterterms as defined by the subtraction formalism adopted 
for the pure-NLO computation~\cite{Frixione:1993dg}. 
The matrix elements have been taken 
from ref.~\cite{Frixione:1993dg}, which we have subsequently matched to 
fortran HERWIG~\cite{Marchesini:1992ch,Corcella:2001bw,Corcella:2002jc},
according to the MC@NLO formalism as described in the following section.

\section{MC@NLO for $Q\Qb$ photoproduction\label{sec:sub}}
The MC@NLO formalism has been introduced in ref.~\cite{Frixione:2002ik},
and applied since then to a fairly large number of hadroproduction processes.
The relevant technical details can be easily found in the literature,
and we shall therefore refrain from giving them again here.
We limit ourselves to recall that in the context of MC@NLO the matching 
of an NLO computation with a PSMC requires 
one to modify the short-distance cross sections that enter the former,
with the inclusion of the so-called Monte Carlo (MC) subtraction terms,
that are responsible for removing any double counting at the NLO.
In turn, the MC subtraction terms have a factorized structure, 
being essentially constructed with Born-level cross sections 
and with process-independent branching kernels, which can be computed
once and for all once the PSMC is chosen that will be used
in the shower phase.

In order to determine the right combination of the Born matrix elements
and branching kernels that enter the MC subtraction terms for a given 
production process, one formally expands the all-order PSMC cross
sections to NLO. The way in which the results so
obtained are manipulated to construct the MC subtraction
terms used in computer programmes is straightforward, and has been
described in several papers. Here, we only give the results
for the perturbative expansion mentioned above. Given the similarity
between heavy-quark photo- and hadroproduction, we adopt the same
notation as in ref.~\cite{Frixione:2003ei}, where the latter process
was studied. The PSMC cross sections read:
\beq
d\sigma\xMCB=\sum_{b}\sum_{L}\sum_{l}d\sigma_{eb}^{(L,l)}\xMCB\,,
\label{MCatas}
\eeq
where the first sum in eq.~(\ref{MCatas}) runs over real-emission
parton processes, eqs.~(\ref{procRpg}) and~(\ref{procRpq}).
The index $L$ runs over the emitting legs and it assumes the values
$+$, $-$, $Q$, and $\Qb$. The index $l$ runs over the  colour structures, 
and it can take the values $t$ and $u$ (clearly, there is no $s$-channel 
colour connection in the case of photoproduction).
Following ref.~\cite{Frixione:2003ei}, we obtain
\beqn
d\sigma_{eb}^{(+,l)}\xMCB &=&
\frac{1}{z_+^{(l)}}
f_{\gamma}^{\Hone}(\bx_{1i}/z_+^{(l)})f_b^{\Htwo}(\bx_{2i})\,
d\hat{\sigma}_{\gamma g}^{(+,l)}\xMCB d\bx_{1i}\,d\bx_{2i}\,,
\label{eq:spl}
\\
d\sigma_{eb}^{(-,l)}\xMCB &=&
\frac{1}{z_-^{(l)}}
f_{\gamma}^{\Hone}(\bx_{1i})f_b^{\Htwo}(\bx_{2i}/z_-^{(l)})\,
d\hat{\sigma}_{\gamma g}^{(-,l)}\xMCB d\bx_{1i}\,d\bx_{2i}\,,
\label{eq:smn}
\\
d\sigma_{eg}^{(Q,l)}\xMCB &=&
f_{\gamma}^{\Hone}(\bx_{1f})f_b^{\Htwo}(\bx_{2f})\,
d\hat{\sigma}_{\gamma g}^{(Q,l)}\xMCB d\bx_{1f}\,d\bx_{2f}\,,
\label{eq:sQ}
\\
d\sigma_{eg}^{(\Qb,l)}\xMCB &=&
f_{\gamma}^{\Hone}(\bx_{1f})f_b^{\Htwo}(\bx_{2f})\,
d\hat{\sigma}_{\gamma g}^{(\Qb,l)}\xMCB d\bx_{1f}\,d\bx_{2f}\,.
\label{eq:sQb}
\eeqn
These equations are identical to eqs.~(5.2)--(5.5) of 
ref.~\cite{Frixione:2003ei}, apart from the obvious notational changes
due to the incoming electron in place of a hadron. The short distance
cross sections in eqs.~(\ref{eq:spl})--(\ref{eq:sQb}) are:

\noindent
$\bullet$~$\gamma g$~initial state ($l=t,u$)
\beqn
d\hat\sigma_{\gamma g}^{(-,l)}\xMCB &=& \frac{\as}{4\pi}\,
\frac{d\xi_-^{(l)}}{\xi_-^{(l)}}dz_-^{(l)}
P^{(0)}_{gg}(z_-^{(l)})\,\dsb_{\gamma g}
\stepf\left((z_-^{(l)})^2-\xi_-^{(l)}\right),
\label{spgptwo}
\\
d\hat\sigma_{\gamma g}^{(Q,l)}\xMCB &=& \frac{\as}{2\pi}\,
\frac{d\xi_Q^{(l)}}{\xi_Q^{(l)}}dz_Q^{(l)}
P^{(0)}_{qq}(z_Q^{(l)})\,\dsb_{\gamma g}^{(l)}
\stepf\left(1-\xi_Q^{(l)}\right)
\stepf\left((z_Q^{(l)})^2-\frac{2m^2}{|\bt_Q|\xi_Q^{(l)}}\right), \nonumber \\
~\\
d\hat\sigma_{\gamma g}^{(\Qb,l)}\xMCB &=& d\hat\sigma_{\gamma g}^{(Q,l)}\xMCB
\left(\bt_Q\to\bt_{\Qb}, z_Q^{(l)}\to z_{\Qb}^{(l)},
\xi_Q^{(l)}\to \xi_{\Qb}^{(l)}\right).
\eeqn

\noindent
$\bullet$~$\gamma q$~initial state
\beqn
d\hat\sigma_{\gamma q}^{(+,u)}\xMCB &=& \frac{\aem}{2\pi}\,
\frac{d\xi_+^{(u)}}{\xi_+^{(u)}}dz_+^{(u)}
P^{(0)}_{q\gamma}(z_+^{(u)})\,\dsb_{\bar{q}q}
\stepf\left((z_+^{(u)})^2-\xi_+^{(u)}\right),
\label{spqpone}
\\
d\hat\sigma_{\gamma q}^{(-,t)}\xMCB &=& \frac{\as}{2\pi}\,
\frac{d\xi_-^{(t)}}{\xi_-^{(t)}}dz_-^{(t)}
P^{(0)}_{gq}(z_-^{(t)})\,\dsb_{\gamma g}
\stepf\left((z_-^{(t)})^2-\xi_-^{(t)}\right).
\label{spqptwo}
\eeqn
$\bullet$~$\gamma \bar q$~initial state
\beqn
d\hat\sigma_{\gamma \bar q}^{(+,t)}\xMCB &=& \frac{\aem}{2\pi}\,
\frac{d\xi_+^{(t)}}{\xi_+^{(t)}}dz_+^{(t)}
P^{(0)}_{q\gamma}(z_+^{(t)})\,\dsb_{\bar{q}q}
\stepf\left((z_+^{(t)})^2-\xi_+^{(t)}\right),
\label{spqbarpone}
\\
d\hat\sigma_{\gamma \bar q}^{(-,u)}\xMCB &=& \frac{\as}{2\pi}\,
\frac{d\xi_-^{(u)}}{\xi_-^{(u)}}dz_-^{(u)}
P^{(0)}_{gq}(z_-^{(u)})\,\dsb_{\gamma g}
\stepf\left((z_-^{(u)})^2-\xi_-^{(u)}\right).
\label{spqbarptwo}
\eeqn
The Born cross sections $\dsb_{\gamma g}$, $\dsb_{\bar{q}q}$ or
$\dsb_{\gamma g}^{(l)}$ have to be computed using the relevant definitions of 
$\bs$, $\bt$ and $\bu$, as explained in ref.~\cite{Frixione:2003ei}.
One also defines~\cite{Odagiri:1998ep}
\beq\label{eq:dsbtu}
\dsb_{\gamma g}^{(t)} = \dsb_{\gamma g}\frac{\bu/\bt}{\bu/\bt+\bt/\bu}
 =  \frac{\dsb_{\gamma g}}{1+\bt^2/\bu^2}\;,\;\;\;
\dsb_{\gamma g}^{(u)} = \frac{\dsb_{\gamma g}}{1+\bu^2/\bt^2}\;.
\eeq
We stress that eqs.~(\ref{spqpone}) and (\ref{spqbarpone}) describe 
QED branchings. When inserted in the MC@NLO short-distance cross 
sections, they subtract an MC contribution which is generated when 
the hadronic-photon component is showered.

The situation of the contributions to the NLO expansion of PSMC cross
sections is summarized in table~\ref{tab:proc}.
%%%%%%%%%%%%%%%%%%%%%%%%%%%%%%%%%%%%%%%%%%%%%%%%%%%%%%%%%%%%%%%%%%%%%%%%%%
\begin{table}[htb]
\renewcommand{\arraystretch}{1.2}
\begin{center}
\begin{tabular}{|c||c|c|c|}
\hline
                & $q\bar{q}\to Q\Qb$ & $\bar{q}q\to Q\Qb$ & $\gamma g\to Q\Qb$ \\\hline\hline
$\gamma g$      &                    &                    & $-(t, u)$, $Q(t,u)$, $\Qb(t,u)$ \\\hline
$\gamma q$      &                    & $+(u)$             & $-(t)$ \\\hline
$\gamma\bar{q}$ & $+(t)$             &                    & $-(u)$ \\\hline
\end{tabular}
\end{center}
\caption{\label{tab:proc}
Short-distance contributions to MC subtraction terms, from Born processes
$q\bar{q}\to Q\Qb$, $\bar{q}q\to Q\Qb$, and $\gamma g\to Q\Qb$. The three
rows correspond to real-emission processes, identified by their incoming
partons. Each entry lists the emitting legs (+, --, $Q$, $\Qb$); for each 
emitting leg, we report in parentheses the different contributions $l$,
according to the possible colour flows (which correspond to 
choosing $E_0^2=|\bl|/2$).
}
\end{table}
%%%%%%%%%%%%%%%%%%%%%%%%%%%%%%%%%%%%%%%%%%%%%%%%%%%%%%%%%%%%%%%%%%%%%%%%%%

\section{Comparisons to measurements of heavy-quark production
 at HERA\label{sec:meas}}
In this section the MC@NLO predictions will be compared to selected 
measurements of various
observables relevant to the production of $B$ and of $D^{*\pm}$ mesons,
as reported by the H1 and ZEUS experiments.
The quark masses and PDFs used by MC@NLO are given in 
table~\ref{tab:parameters}. 
The uncertainty band in MC@NLO is computed by varying the factorization 
and renormalization scales independently, by a factor of two up and 
down from the default scale $1/2(m_{Q,T}+m_{\bar{Q},T})$, and
by taking the envelope of the results obtained in this way.
For charm quarks these scales may become fairly small, and indeed 
charm production is pushing the applicability of perturbative QCD 
to its limits -- this is one of the reasons why it is interesting
to compare MC@NLO results for this process to data.

%%%%%%%%%%%%%%%%%%%%%%%%%%%%%%%%%%%%%%%%%%%%%%%%%%%%%%%%%%%%%%%%%%%%%%%%%%
\begin{table}
  \begin{center}
    \begin{tabular}{|c|c|}\hline
      $m_b$ & 4.75 GeV \\\hline
      $m_c$ & 1.5  GeV \\\hline
      Proton PDF & Cteq6.6   \\\hline
      Photon PDF & GRV        \\\hline
    \end{tabular}
  \end{center}
  \caption{Parameter settings used by MC@NLO.}
\label{tab:parameters}
\end{table}
%%%%%%%%%%%%%%%%%%%%%%%%%%%%%%%%%%%%%%%%%%%%%%%%%%%%%%%%%%%%%%%%%%%%%%%%%%

In order to quantify the level of agreement between
MC@NLO predictions and data, the quantity $\chi^2/ndf$ has been
calculated for each data set, by taking into account both
theory and experimental uncertainties.  
The relevant results are summarized in table~\ref{tab:bchi} for 
bottom measurements, and in table~\ref{tab:dstarchi} for 
charm measurements.

No effort has been made here to tune HERWIG (version 6.510) 
to HERA data, since the idea is that of making an
out-of-the-box comparison. The only exception to this rule is an
{\em overall} rescaling, applied to all hadron-level observables in order
to have values of branching ratios consistent with those reported
by the PDG~\cite{Nakamura:2010zzi} -- these rescaling factors
are equal to $1.34$ and $1.5$ for charm and bottom respectively.

\subsection{$B$-hadron production}
Bottom-flavoured hadrons are typically tagged by searching for the
muons that arise from $W$'s, which in turn come from the weak decay
of the lowest-lying $b$-hadrons into lighter-quark states. These
muons will in general have a large momentum transverse to the jet axis, so
called $p_T^{\rm rel}$.  Also, the vertex from which these muons are radiated
will be displaced relative to the hard interaction of the event, and this
displacement is proportional to the lifetime of the $b$-hadron. Often, only the
transverse component $\delta$ of this displacement is used in the $b$-tagging.
These two methods of tagging the $b$-quarks may be combined to further
enhance the signal.

In this section, comparisons will be made with three measurements performed
by H1 and ZEUS at HERA. They are: 
\begin{enumerate}
\item ``Measurement of beauty photoproduction using decays into muons in dijet
  events at HERA'', by the ZEUS collaboration~\cite{Chekanov:2008tx};
\item  ``Measurement of beauty production at HERA using events with muons and
  jets'', by the H1 collaboration~\cite{Aktas:2005zc};
\item ``Beauty photoproduction measured using decays into muons in 
  dijet events in $ep$ collisions at $\sqrt{s} = 318$~GeV'', by the 
  ZEUS collaboration~\cite{Chekanov:2003si}. 
\end{enumerate}
These will be referred to as ZEUS-09, H1-05 and ZEUS-03 respectively. The
first two analyses use 
the combined method of both $p_T^{\rm rel}$ and $\delta$ in the
tagging of the $b$-quarks, while in ZEUS-03 only $p_T^{\rm rel}$ is used. The
experimental cuts made for the bottom analyses are summarized in table
\ref{tab:bcuts}. These cuts result in the visible cross-sections listed in
table \ref{tab:bvisible}, where also the MC@NLO predictions are shown;
theory and data are in good agreement.

%%%%%%%%%%%%%%%%%%%%%%%%%%%%%%%%%%%%%%%%%%%%%%%%%%%%%%%%%%%%%%%%%%%%%%%%%%
\begin{table}
  \begin{center}
    \begin{tabular}{|c|c|c|c|c|}\hline
      Analysis               &  ZEUS-09          & H1-05           & ZEUS-03           \\\hline
      $\sqrt{s}$             & 318 GeV           & 318 GeV         & 318 GeV          \\\hline 
      $Q^2$                  & $<1 {\rm ~GeV}^2$ & $<1 {\rm ~GeV}^2$& $<1 {\rm ~GeV}^2$\\\hline
      $y_{JB}$               & $0.2-0.8$         & $0.2-0.8$        & $0.2-0.8$         \\\hline
      $p_{t}(\mu)$           & $ > 2.5$~GeV      & $ > 2.5$~GeV     & $ > 2.5$~GeV     \\\hline
      $\eta(\mu)$            & -$1.6-1.3$        & -$0.55-1.1$     & -$1.48-2.3$       \\\hline
      $p_T({\rm jet}_{1,2})$ & $7,6$~GeV         & $7,6$~GeV       & $7,6$~GeV            \\\hline  
      $\eta$(jet)           & -$2.5-2.5$        & -$2.5-2.5$       & -$2.5-2.5$        \\\hline
    \end{tabular}
  \end{center}
  \caption{A summary of the cuts relevant to the bottom analyses considered
           in this paper.}
  \label{tab:bcuts}
\end{table}
%%%%%%%%%%%%%%%%%%%%%%%%%%%%%%%%%%%%%%%%%%%%%%%%%%%%%%%%%%%%%%%%%%%%%%%%%%
%%%%%%%%%%%%%%%%%%%%%%%%%%%%%%%%%%%%%%%%%%%%%%%%%%%%%%%%%%%%%%%%%%%%%%%%%%
\begin{table}
  \begin{center}
    \begin{tabular}{|c|c|c|c|c|}\hline
      Visible $\sigma$ [pb]       &  Measured           &  MC@NLO        
      \\\hline
      ZEUS-09                & $38.6^{+5.78}_{-6.02}$ & $42.08\pm4.91$ \\\hline
      H1-05                  & $38.4\pm6.38$       & $33.71\pm2.89$ \\\hline
      ZEUS-03                & $50.25\pm6.45$      & $48.39\pm3.87$ \\\hline
    \end{tabular}
  \end{center}
  \caption{The visible cross sections within the cuts listed in table
    \ref{tab:bcuts}. The different measurements are compared to MC@NLO
    predictions.} 
  \label{tab:bvisible}
\end{table}
%%%%%%%%%%%%%%%%%%%%%%%%%%%%%%%%%%%%%%%%%%%%%%%%%%%%%%%%%%%%%%%%%%%%%%%%%%
%%%%%%%%%%%%%%%%%%%%%%%%%%%%%%%%%%%%%%%%%%%%%%%%%%%%%%%%%%%%%%%%%%%%%%%%%%
\begin{table}
  \begin{center}
    \begin{tabular}{|c|c|c|c|}\hline
      ZEUS-09 & MC@NLO 
      \\\hline
      $p_T(\mu)$ & 0.18 \\\hline
      $\eta(\mu)$ & 0.05 \\\hline
      $x_\gamma({\rm jets})$ & 0.59 \\\hline
      $\Delta\phi({\rm jets})$ & 1.22 \\\hline
      $\Delta\phi({\rm jets})|x_\gamma^{\rm obs}<0.75$ & 0.52 \\\hline\hline
      H1-05 & MC@NLO 
      \\\hline
      $p_T(\mu)$ & 0.89 \\\hline
      $\eta(\mu)$ & 0.11 \\\hline
      $x_\gamma({\rm jets})$ & 0.48 \\\hline\hline
      ZEUS-03 & MC@NLO 
      \\\hline
      $p_T(\mu)$ & 0.78 \\\hline
      $\eta(\mu)$ & 0.34 \\\hline
      $p_T(b-{\rm jet})$ & 0.09 \\\hline
      $p_T(b)$ & 0.65 \\\hline
      $x_\gamma({\rm jets})$ & 0.17 \\\hline
    \end{tabular}
  \end{center}
  \caption{The $\chi^2/ndf$ for all distributions in 
    the bottom measurements shown in this paper.}
  \label{tab:bchi}
\end{table}
%%%%%%%%%%%%%%%%%%%%%%%%%%%%%%%%%%%%%%%%%%%%%%%%%%%%%%%%%%%%%%%%%%%%%%%%%%

\begin{figure}
  \begin{center}
  \includegraphics[width=0.4\columnwidth]{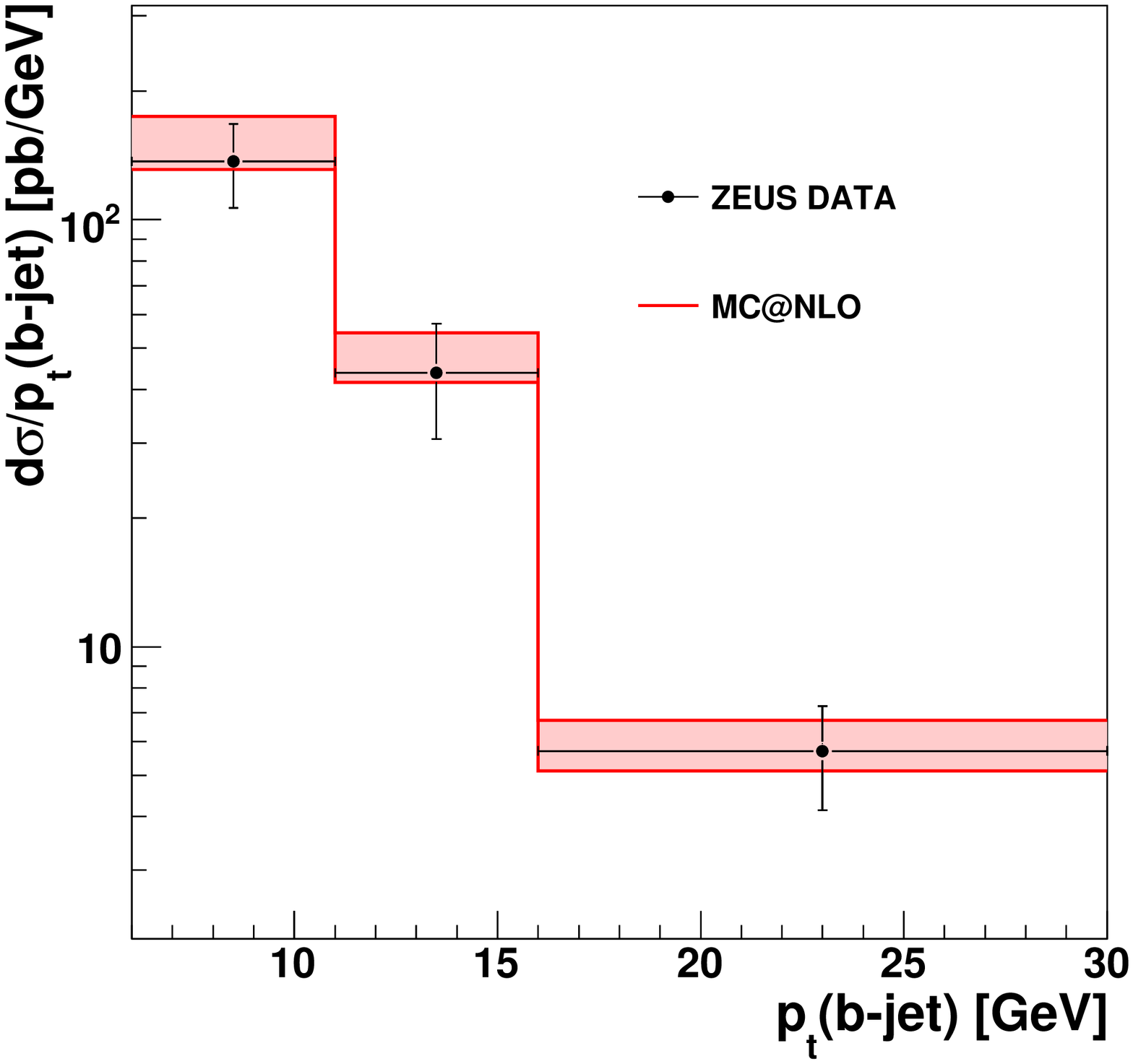}
  \includegraphics[width=0.4\columnwidth]{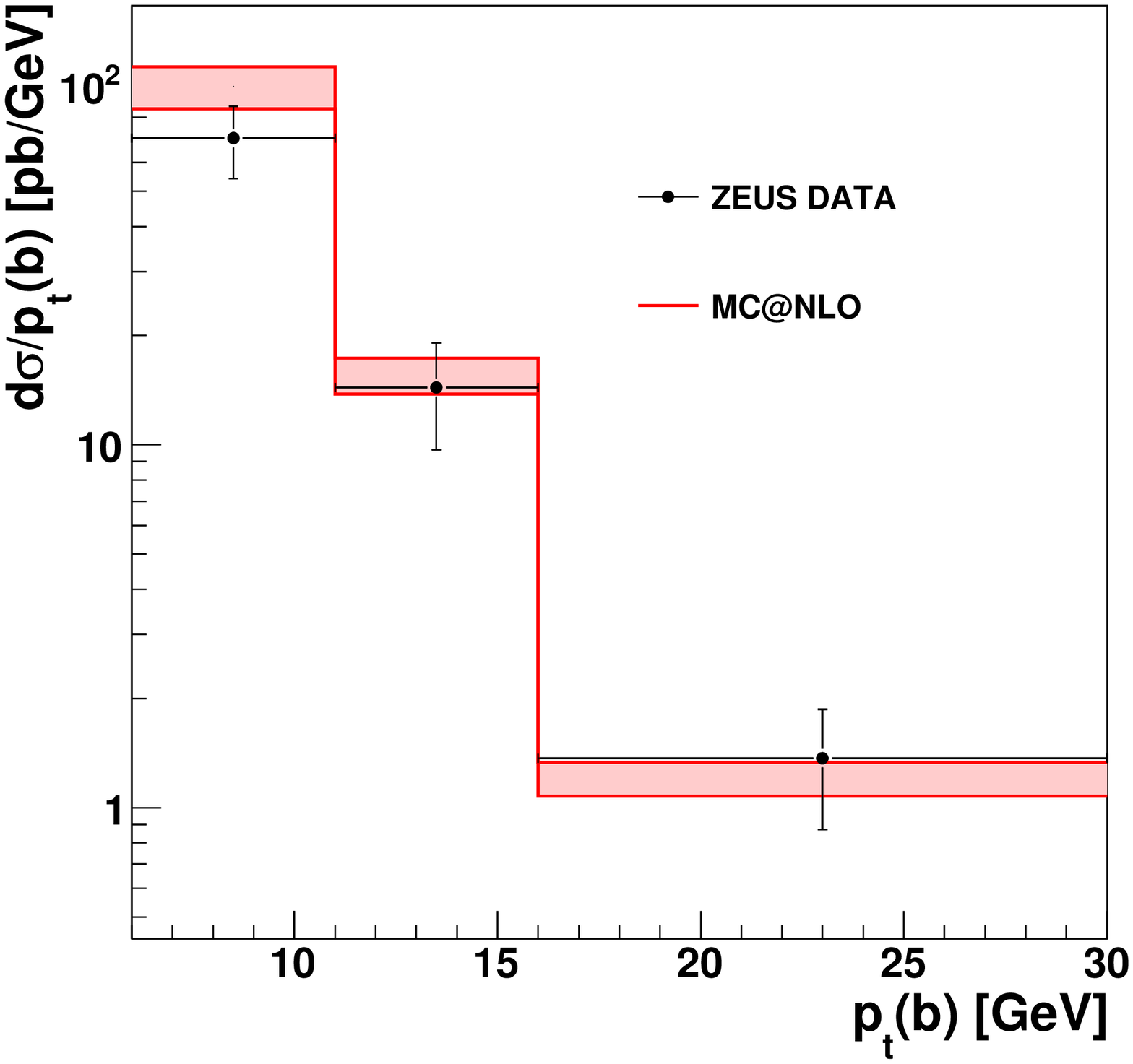}
  \end{center}
  \caption{Distributions of $p_T(b-{\rm jet})$ and $p_T(b)$ from the measurement
    ZEUS-03. The MC@NLO band includes full independent scale variations.}
  \label{fig:bb}
\end{figure}

In fig.~\ref{fig:bb} the $p_T$ of the jet containing a $b$-quark is
shown, as measured in ZEUS-03. This spectrum is then used by the ZEUS
collaboration to reconstruct the $p_T$ spectrum of the $b$-quarks, 
using the NLO calculation FMNR~\cite{Frixione:1993dg}.
Both distributions are well described by MC@NLO. The $p_T(b-jet)$ 
prediction of MC@NLO, being at the hadron level, has been rescaled 
for the overall branching ratio factor as discussed before.
On the other hand, the $p_T(b)$ spectrum, being a quantity at the parton
level, has not been rescaled. It should be stressed that the use
of NLO computations matched to parton showers implies that the deconvolution
of data from hadron to parton level is not necessary for
a fair comparison with theoretical predictions; in fact, such
deconvolutions have to be deprecated, since they introduce 
unnecessary theoretical biases in the measurements (e.g., in the
present case, the underlying matrix elements of FMNR and MC@NLO
are the same; hence, the comparison done for $p_T(b)$ is not as significant
as that for $p_T(b-jet)$).
%%%%%%%%%%%%%%%%%%%%%%%%%%%%%%%%%%%%%%%%%%%%%%%%%%%%%%%%%%%%%%%%%%%%%%%%%%
%%%%%%%%%%%%%%%%%%%%%%%%%%%%%%%%%%%%%%%%%%%%%%%%%%%%%%%%%%%%%%%%%%%%%%%%%%

\begin{figure}
  \begin{center}
  \includegraphics[width=0.4\columnwidth]{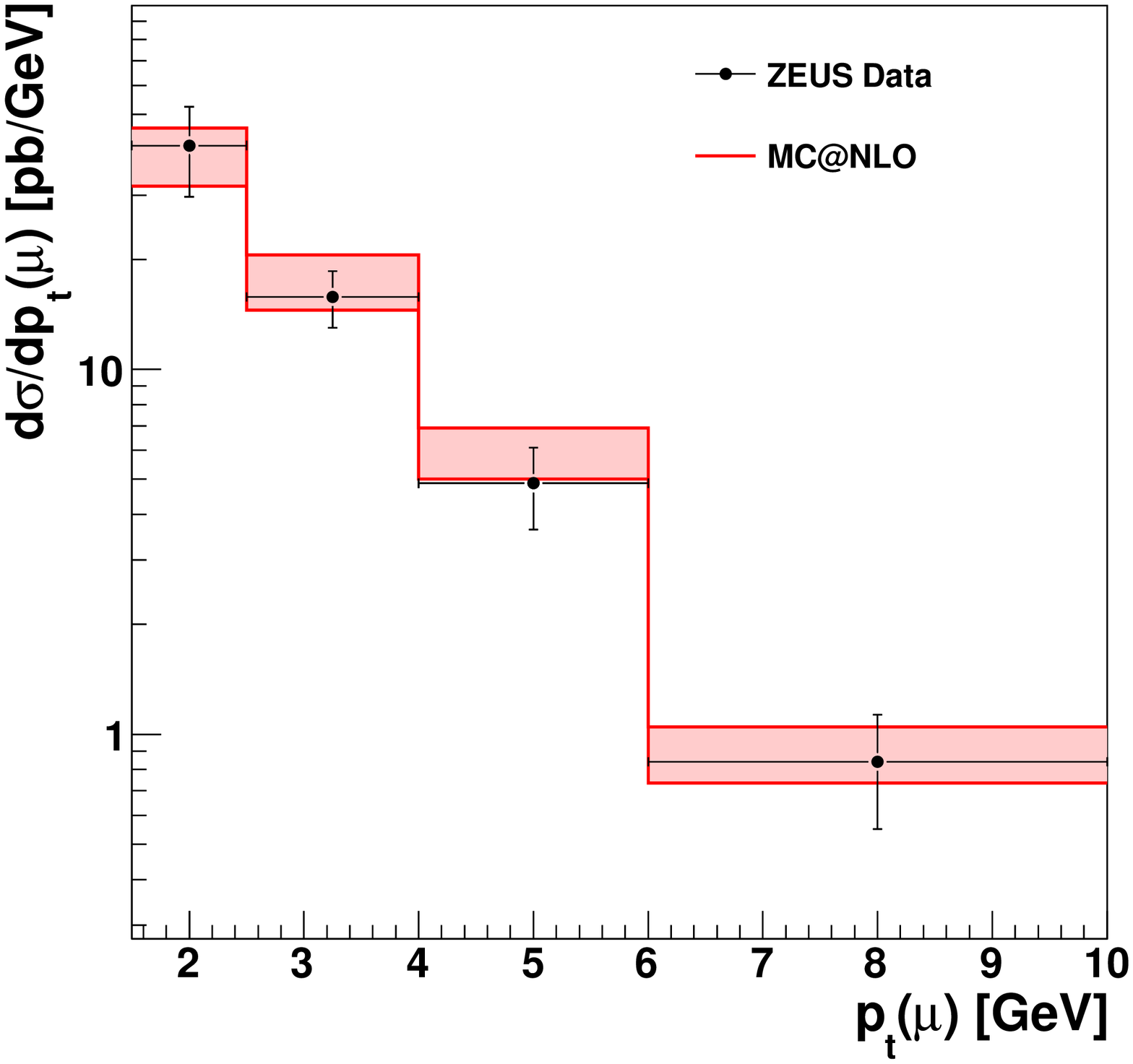}
  \includegraphics[width=0.4\columnwidth]{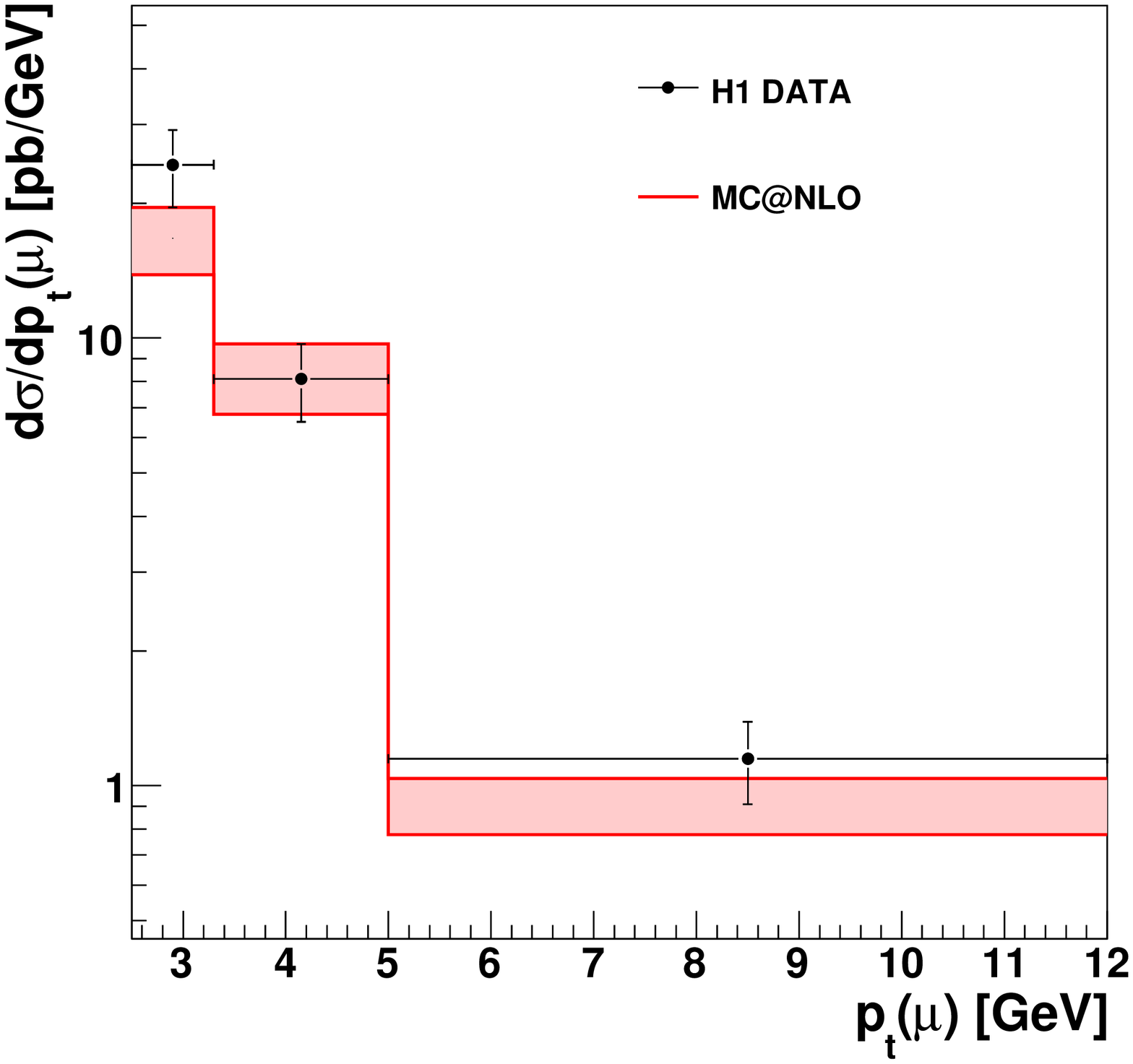}

  \includegraphics[width=0.4\columnwidth]{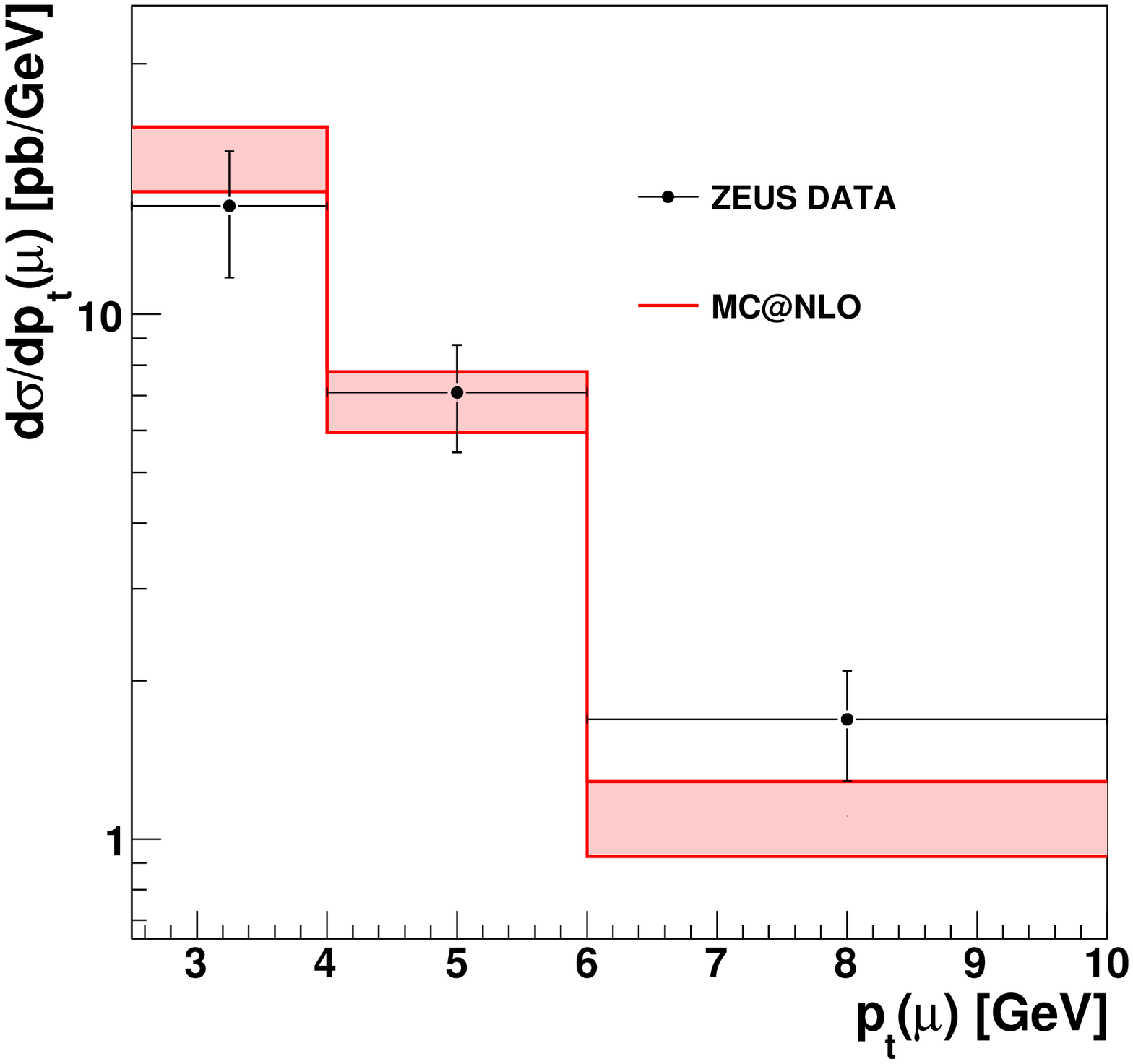}
  \includegraphics[width=0.4\columnwidth]{void.eps}
  \end{center}
  \caption{Distributions of $p_T(\mu)$ from the measurements ZEUS-09 (upper left),
  H1-05 (upper right) and ZEUS-03 (bottom). The MC@NLO band includes full 
  independent scale variations.}
  \label{fig:bpt}
\end{figure}

\begin{figure}
  \begin{center}
  \includegraphics[width=0.4\columnwidth]{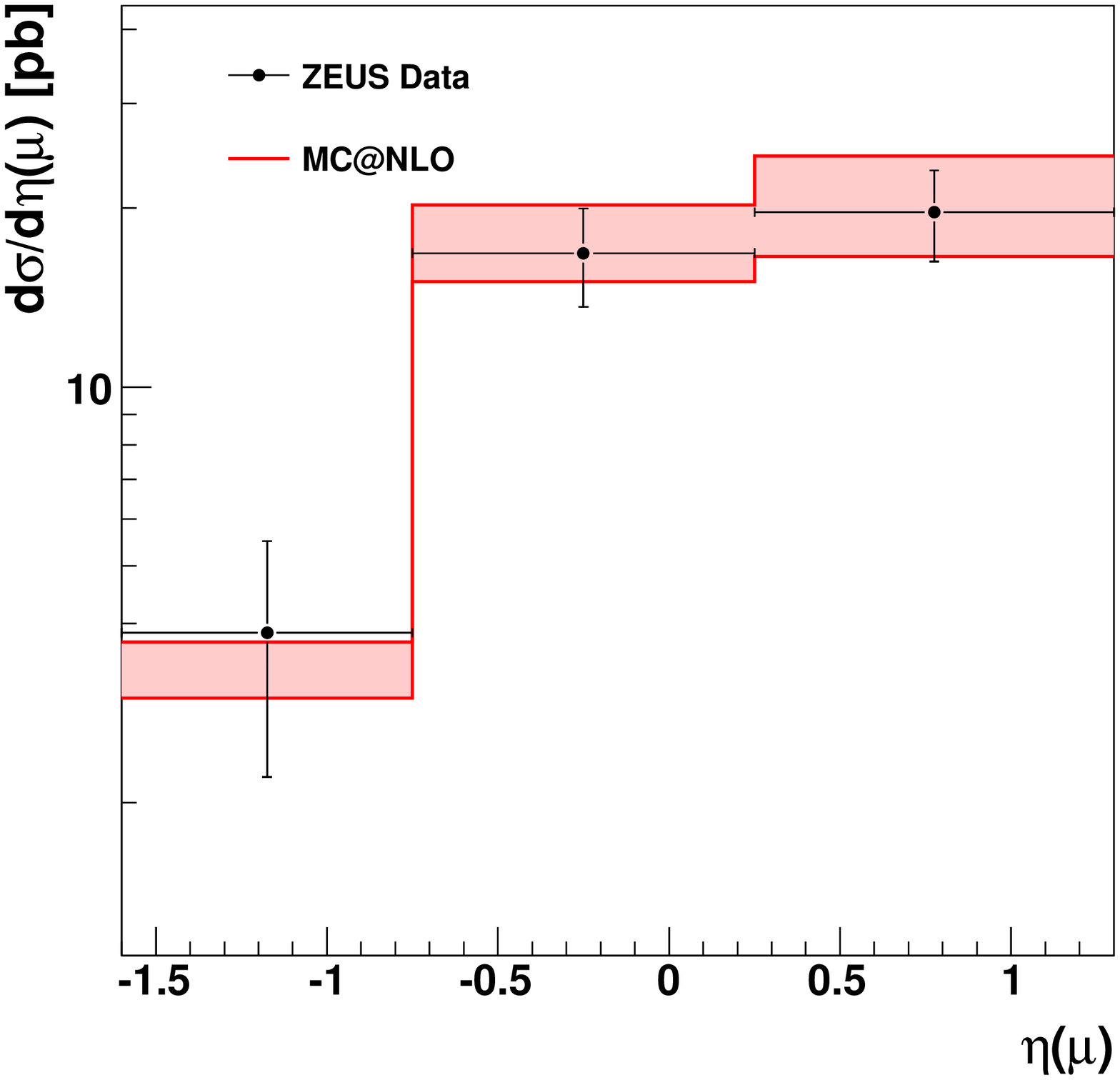}
  \includegraphics[width=0.4\columnwidth]{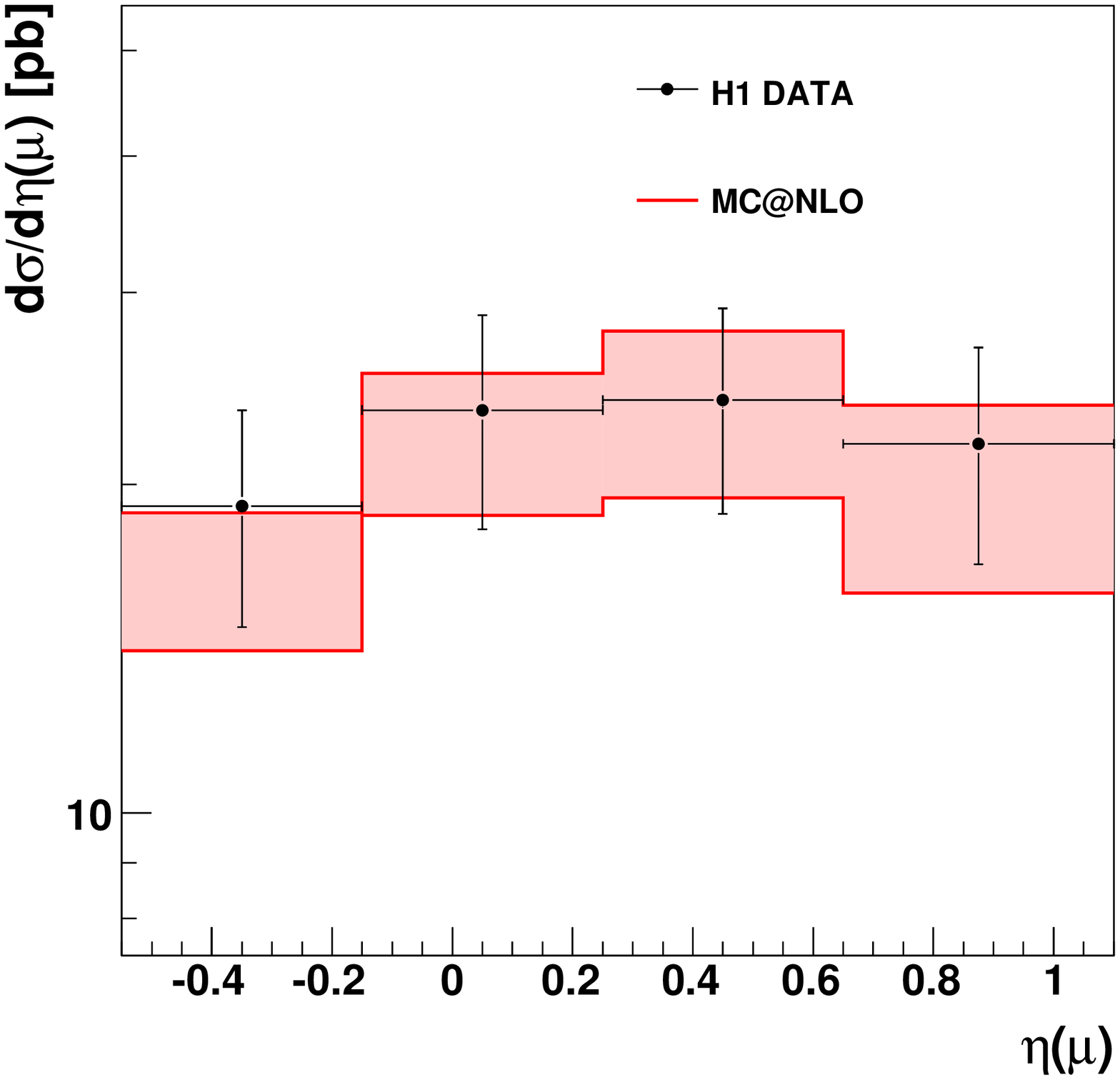}

  \includegraphics[width=0.4\columnwidth]{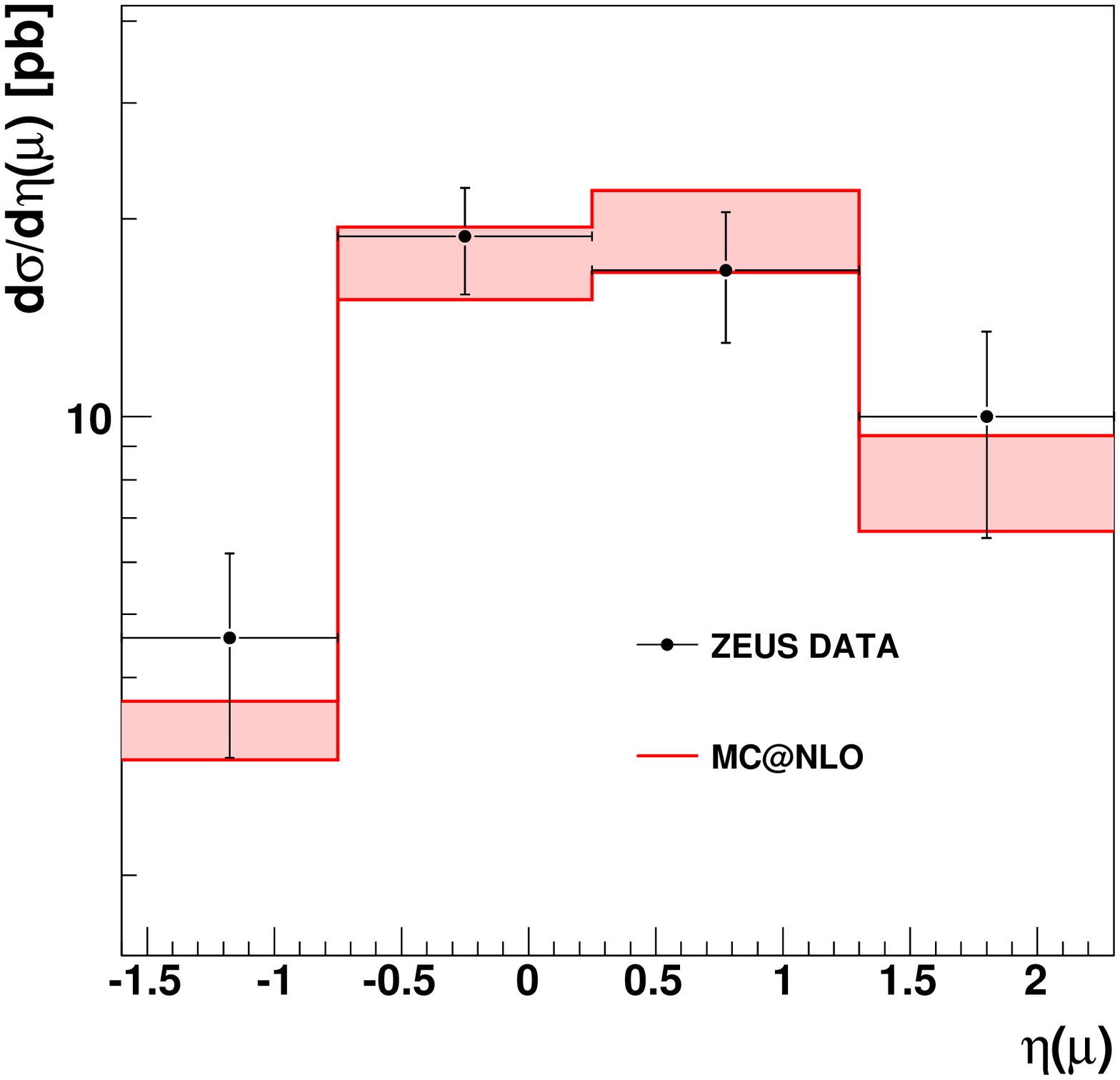}
  \includegraphics[width=0.4\columnwidth]{void.eps}
  \end{center}
  \caption{Distributions of $\eta(\mu)$ from the measurements ZEUS-09 
  (upper left), H1-05 (upper right) and ZEUS-03 (bottom). The MC@NLO 
  band is obtained with independent scale variations.}
  \label{fig:beta}
\end{figure}

In fig.~\ref{fig:bpt} the transverse momentum spectra of the tagged muons
are shown. MC@NLO is describing all three data sets well for this observable.
Also, the scale variations in MC@NLO are at the same level or smaller than the
experimental uncertainties. The rapidity distributions of the muons
are also well described by MC@NLO in all three data sets, as seen in
fig.~\ref{fig:beta}.
%%%%%%%%%%%%%%%%%%%%%%%%%%%%%%%%%%%%%%%%%%%%%%%%%%%%%%%%%%%%%%%%%%%%%%%%%%
%%%%%%%%%%%%%%%%%%%%%%%%%%%%%%%%%%%%%%%%%%%%%%%%%%%%%%%%%%%%%%%%%%%%%%%%%%
%%%%%%%%%%%%%%%%%%%%%%%%%%%%%%%%%%%%%%%%%%%%%%%%%%%%%%%%%%%%%%%%%%%%%%%%%%
%%%%%%%%%%%%%%%%%%%%%%%%%%%%%%%%%%%%%%%%%%%%%%%%%%%%%%%%%%%%%%%%%%%%%%%%%%

\begin{figure}
  \begin{center}
  \includegraphics[width=0.4\columnwidth]{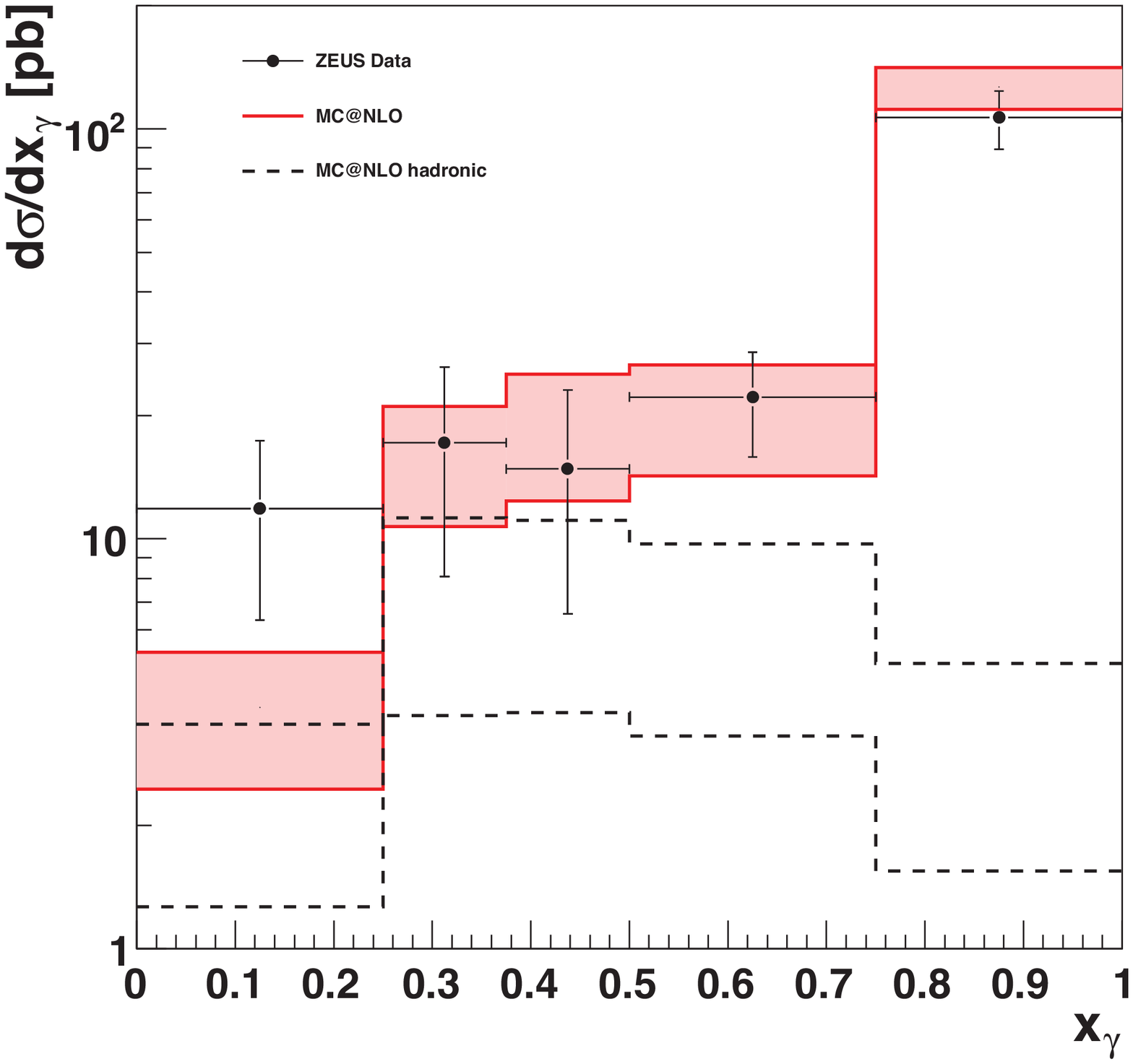}
  \includegraphics[width=0.4\columnwidth]{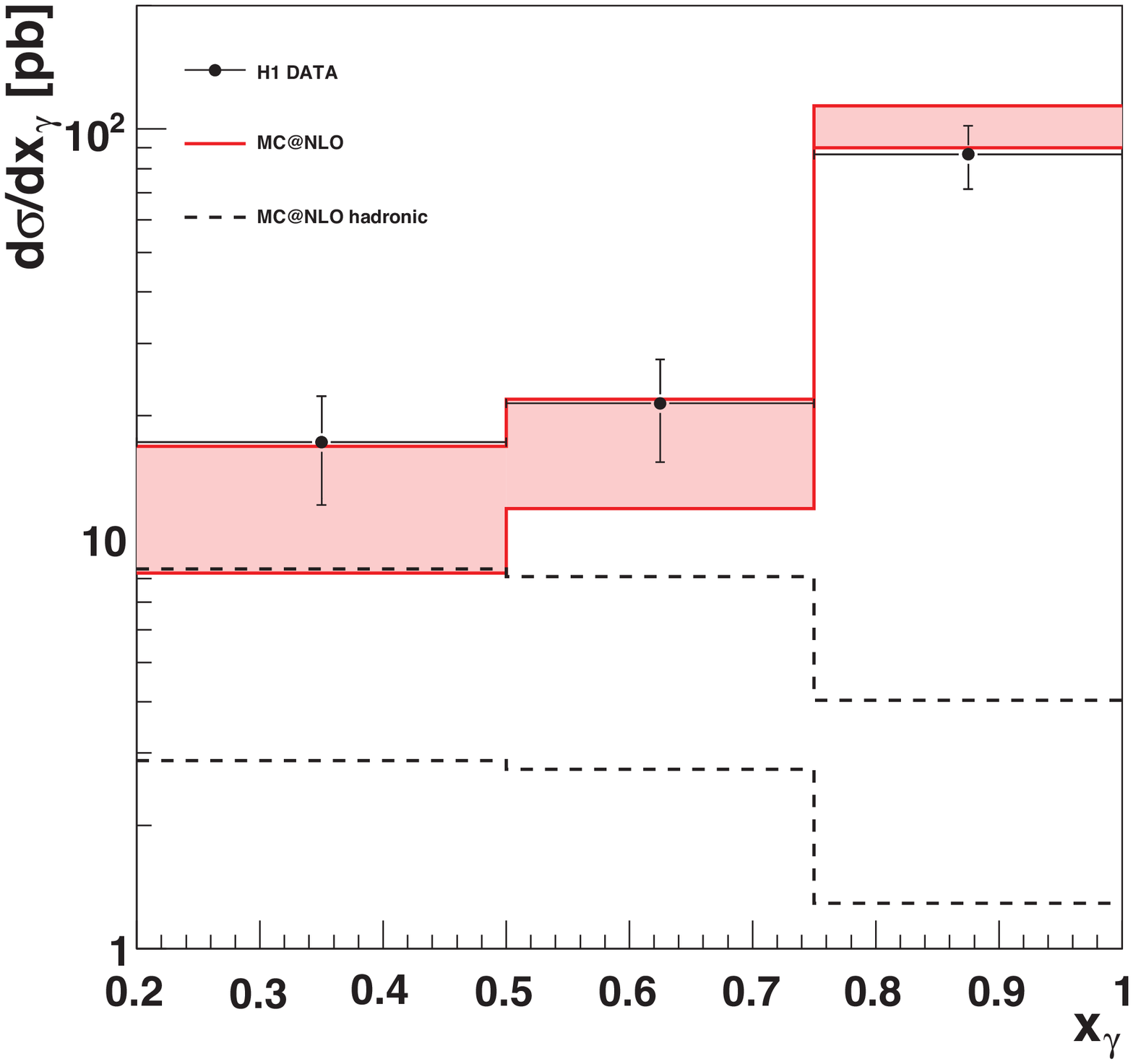}

  \includegraphics[width=0.4\columnwidth]{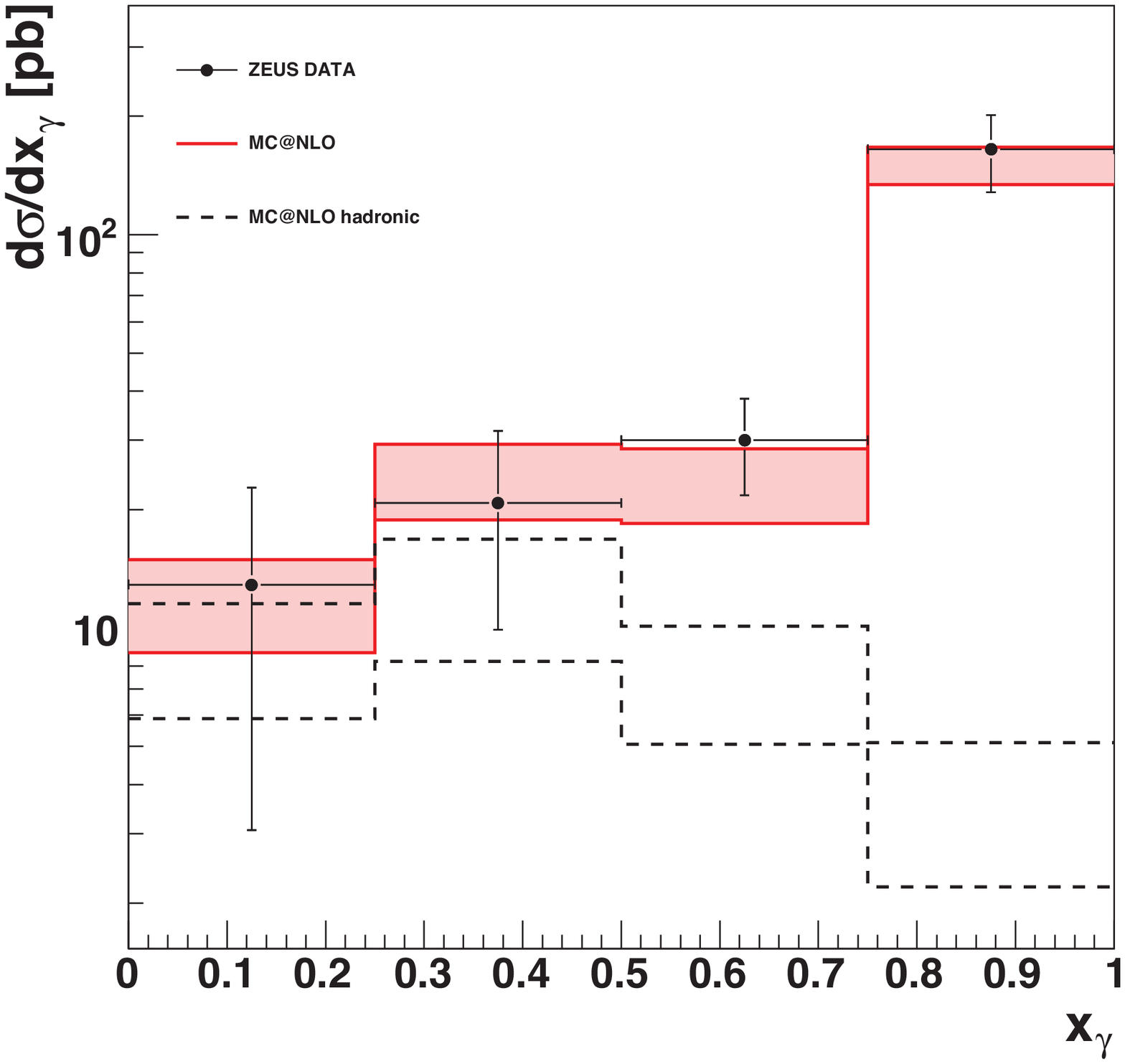}
  \includegraphics[width=0.4\columnwidth]{void.eps}
  \end{center}
  \caption{Distributions of $x_\gamma^{\rm obs}$ from the measurements ZEUS-09 
    (upper left),
    H1-05 (upper right) and ZEUS-03 (bottom). The MC@NLO band includes full 
    independent scale variations.}
  \label{fig:bxgam}
\end{figure}

Two types of correlations have been measured for bottom production. One
is the $x_\gamma^{\rm obs}$ distribution of the two leading jets in 
the measurements, shown in fig.~\ref{fig:bxgam}. The variable
$x_\gamma^{\rm obs}$ is defined by:
\begin{eqnarray}
  x_\gamma^{\rm obs}({\rm jet}_1, {\rm jet}_2)=
  \frac{P_-({\rm jet})_1+P_-({\rm jet}_2)}{\sum_{{\rm All~hadrons}~i}P_-(i)}
\end{eqnarray}
where $P_-=E-P_z$. Thus, $x_\gamma^{\rm obs}$ is the fraction of the measured
hadronic $P_-$ carried by the two leading jets, and at the LO it coincides 
with the fraction of the photon energy which enter into the 
hard interaction. Therefore, the hadronic part of the calculation
is expected to be more important for small values of $x_\gamma^{\rm obs}$.
MC@NLO describes all three measurements well. In these plots, the
hadronic part of the MC@NLO calculation is also shown separate and,
as expected, becomes significant for $x_\gamma^{\rm obs}<0.75$. 
%%%%%%%%%%%%%%%%%%%%%%%%%%%%%%%%%%%%%%%%%%%%%%%%%%%%%%%%%%%%%%%%%%%%%%%%%%
%%%%%%%%%%%%%%%%%%%%%%%%%%%%%%%%%%%%%%%%%%%%%%%%%%%%%%%%%%%%%%%%%%%%%%%%%%

\begin{figure}
  \begin{center}
  \includegraphics[width=0.4\columnwidth]{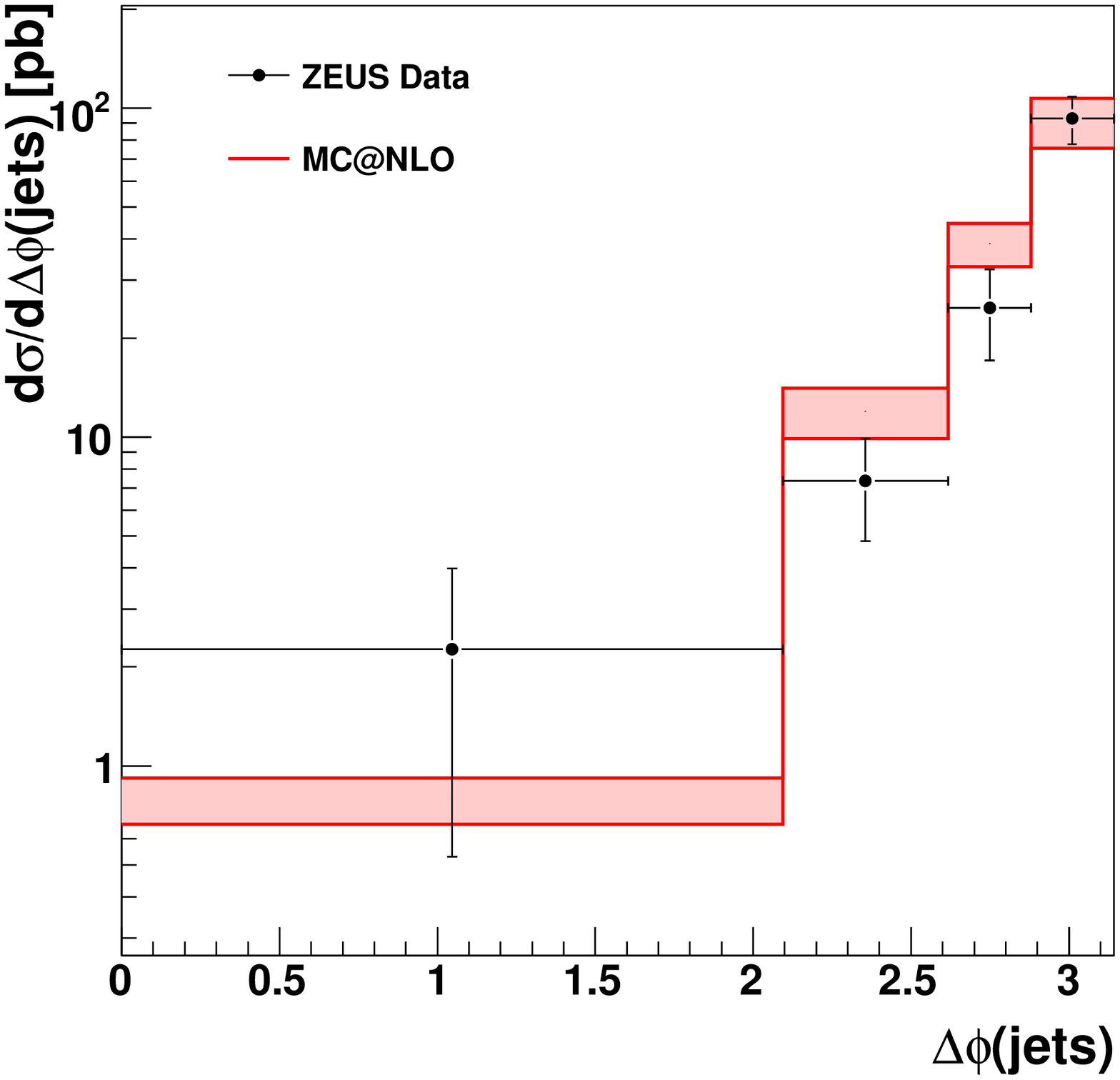}
  \includegraphics[width=0.4\columnwidth]{void.eps}

  \includegraphics[width=0.4\columnwidth]{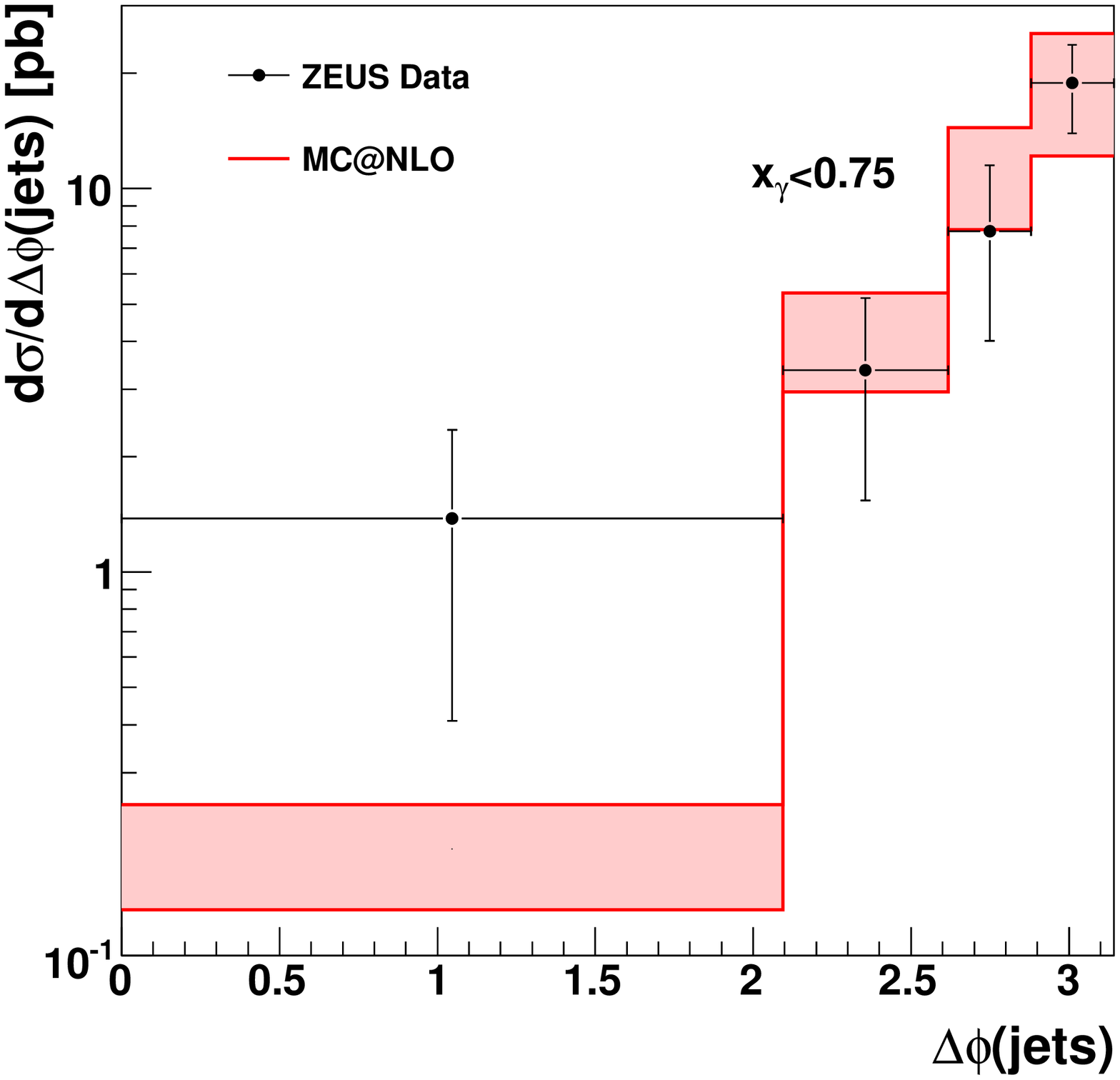}
  \includegraphics[width=0.4\columnwidth]{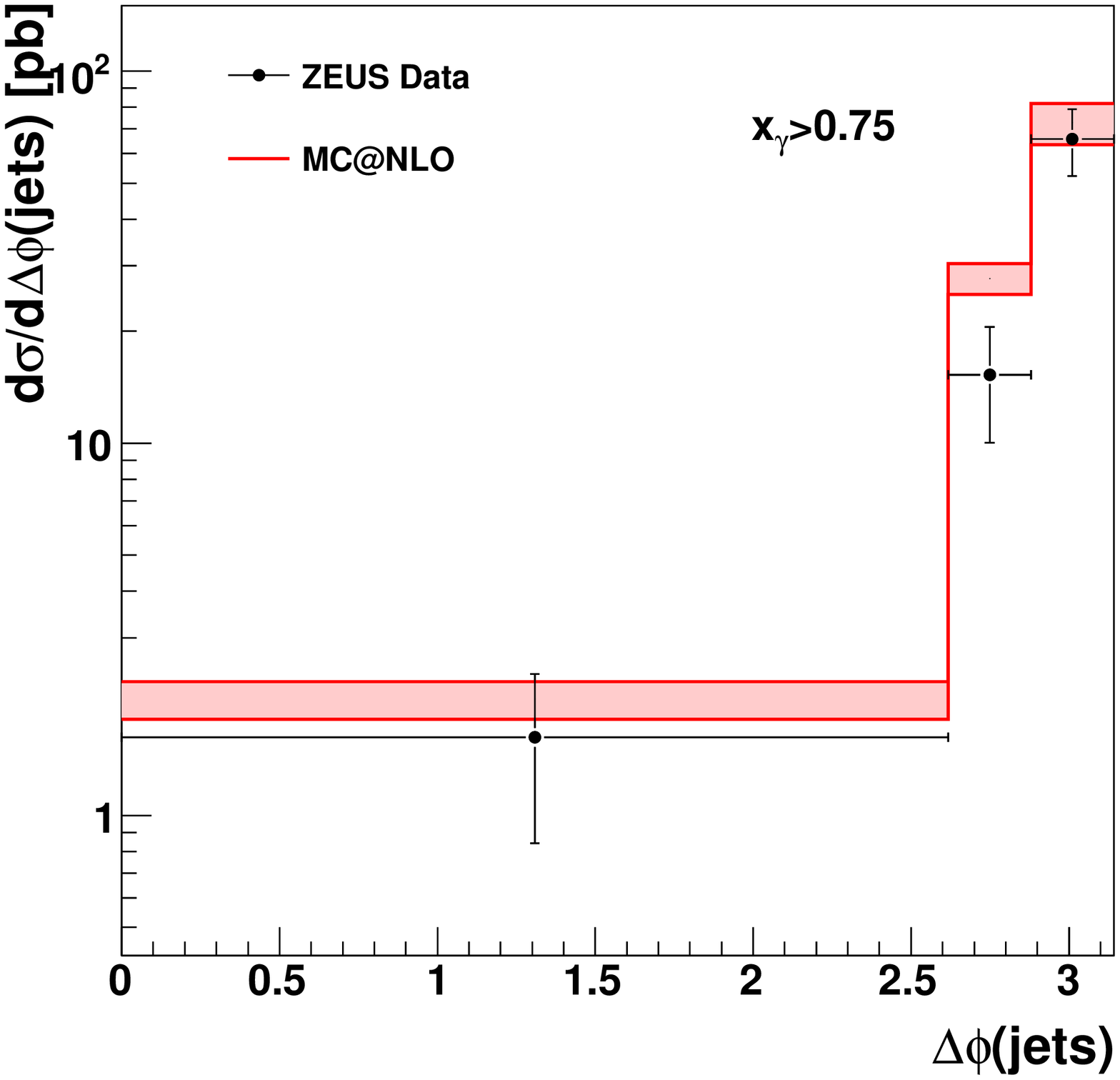}
  \end{center}
  \caption{Distributions of $\Delta\phi({\rm jets})$, unbinned and binned in
    $x_\gamma^{\rm obs}({\rm jets})$ from ZEUS-09. 
    The MC@NLO band includes full independent scale variations.}
  \label{fig:bphi}
\end{figure}

An observable which is highly sensitive to higher-order effects is the
difference in azimuthal angle between the two leading jets. 
At $\Delta\phi\simeq\pi$ one observes the typical logarithmic divergence
of infrared-sensitive variables computed at fixed-order in perturbation
theory; this divergence is suppressed by the Sudakov damping, present
in resummed computations and therefore naturally included in Monte Carlos.
Multiple soft and collinear emissions are also relevant to small 
$\Delta\phi$ values (especially when rather small transverse momenta
are involved), where however there are non-negligible contributions
due to hard matrix elements, present in perturbative computations but
not in ordinary PSMCs. It is therefore clear that MC@NLO, by incorporating
both contributions, is expected to give a better description for this
variable than either fixed-order computations or ordinary PSMCs.
We present the comparison of MC@NLO predictions with data 
in fig.~\ref{fig:bphi}; the agreement is satisfactory, given the
large experimental uncertainties.  This observable is also presented
by separating the large and small $x_\gamma^{\rm obs}$ regions, which
are dominated by the pointlike and hadronic photon components respectively.
The results are also displayed in fig.~\ref{fig:bphi}.
%%%%%%%%%%%%%%%%%%%%%%%%%%%%%%%%%%%%%%%%%%%%%%%%%%%%%%%%%%%%%%%%%%%%%%%%%%
%%%%%%%%%%%%%%%%%%%%%%%%%%%%%%%%%%%%%%%%%%%%%%%%%%%%%%%%%%%%%%%%%%%%%%%%%%

\subsection{$D^{*\pm}$ production}
The $D^{*\pm}$ mesons are detected through the 
so-called golden decay channel:
\begin{eqnarray}
  D^{*\pm}\rightarrow D^0\pi^\pm_{\rm slow}\rightarrow 
 K^\mp\pi^\pm\pi^\pm_{\rm slow}.
\end{eqnarray}
The branching ratio for the golden
decay channel is $\sim 2.6\%$ \cite{Nakamura:2010zzi}, which is comparatively low,  
but the advantage of this channel is that all the final state particles carry 
an electric charge, resulting in three charged tracks in the detectors. 

In this section MC@NLO will be compared to two $D^{*\pm}$ measurements.
These are:
\begin{enumerate}
\item  ``Inclusive $D^*$-Meson Cross Sections and D*-Jet Correlations in 
  Photoproduction at HERA'' by the H1 collaboration~\cite{Aktas:2006ry};
\item ``Inclusive jet cross sections and dijet correlations in $D^{*\pm}$ 
 photoproduction at HERA'' by the ZEUS collaboration~\cite{Chekanov:2005zg}.
\end{enumerate}
These will be referred to as H1-06 and ZEUS-05 respectively.
The experimental cuts applied in the $D^*$ analyses are summarized in table
\ref{tab:ccuts}. These cuts result in the visible cross sections reported
in table~\ref{tab:dstarvisible}, 
together with the theoretical predictions. MC@NLO
describe the H1-06 measurements very well and is two sigma below 
the data for ZEUS-05.

%%%%%%%%%%%%%%%%%%%%%%%%%%%%%%%%%%%%%%%%%%%%%%%%%%%%%%%%%%%%%%%%%%%%%%%%%%
\begin{table}
  \begin{center}
    \begin{tabular}{|c|c|c|c|c|c|}\hline
      Analysis               
      & H1-06  &ZEUS-05
      \\\hline
      $\sqrt{s}$             
      & 318 GeV            & 318 GeV 
      \\\hline 
      $Q^2$                  
      &$<0.01 {\rm ~GeV}^2$&$<1 {\rm ~GeV}^2$
      \\\hline
      $y_{JB}$               
      & $0.29-0.65$        & $0.19-0.87$
      \\\hline
      $p_{t}(D^*)$           
      & $ > 2$~GeV         & $ > 3$~GeV     
      \\\hline
      $\eta(D^*)$            
      & -$1.5-1.5$         & -$1.5-1.5$     
      \\\hline
      $\eta$(jet)           
      & -$1.5-1.5$        & -$1.5-2.4$    
      \\\hline
      $p_T({\rm jet}_{1,2})$ 
      & 4,3 GeV            &              
      \\\hline  
    \end{tabular}
  \end{center}
  \caption{A summary of the cuts in the $D^*$ meson measurements.}
  \label{tab:ccuts}
\end{table}
%%%%%%%%%%%%%%%%%%%%%%%%%%%%%%%%%%%%%%%%%%%%%%%%%%%%%%%%%%%%%%%%%%%%%%%%%%
%%%%%%%%%%%%%%%%%%%%%%%%%%%%%%%%%%%%%%%%%%%%%%%%%%%%%%%%%%%%%%%%%%%%%%%%%%
\begin{table}
  \begin{center}
    \begin{tabular}{|c|c|c|c|c|}\hline
      Visible C-S [nb]       &  Measured       &  MC@NLO  
      \\\hline
      H1-06 inclusive $D^*$  &  $6.45\pm0.83$  &  $6.45\pm0.78$\\\hline
      H1-06 $D^*$+jets       &  $3.01\pm0.44$  &  $2.88\pm0.29$\\\hline
      ZEUS-05                &  $6.80\pm0.26$  &  $5.77\pm0.42$\\\hline
    \end{tabular}
  \end{center}
  \caption{The resulting visible cross-sections from the cuts listed in table
    \ref{tab:ccuts}, for the different measurements as well for the MC@NLO
    predictions. H1-06 consists of two measurements: of inclusive $D^*$ and of
    $D^* +~$jets. These are listed separately here.}
  \label{tab:dstarvisible}
\end{table}
%%%%%%%%%%%%%%%%%%%%%%%%%%%%%%%%%%%%%%%%%%%%%%%%%%%%%%%%%%%%%%%%%%%%%%%%%%
%%%%%%%%%%%%%%%%%%%%%%%%%%%%%%%%%%%%%%%%%%%%%%%%%%%%%%%%%%%%%%%%%%%%%%%%%%
\begin{table}
  \begin{center}
    \begin{tabular}{|c|c|c|c|}\hline
      H1-06 inclusive& MC@NLO $\chi^2/nds$ 
      \\\hline
      $p_T(D^*)$ & 1.41 \\\hline
      $\eta(D^*)$ & 0.46  \\\hline
      $\eta(D^*)\big|{p_T(D^*)>4.5~{\rm GeV}}$ & 1.67 \\\hline\hline
      H1-06 $D^*$+jet&MC@NLO $\chi^2/nds$
      \\\hline
      $p_T(D^*)$ & 2.49 \\\hline
      $\eta(D^*)$& 0.52  \\\hline
      $x_\gamma^{\rm obs}$& 1.94  \\\hline
      $\Delta\phi(D^*, {\rm jet})$& 0.67 \\\hline
      $\Delta\eta(D^*, {\rm jet})$& 0.28 \\\hline\hline
      ZEUS-05 & MC@NLO $\chi^2/nds$
      \\\hline
      $p_T(D^*)$ & 0.94 \\\hline
      $x_\gamma^{\rm obs}$& 1.05 \\\hline
      $\Delta\phi(D^*, {\rm jet})$& 1.37 \\\hline
      $p_T({\rm jj})$ & 1.01 \\\hline
      $M_{\rm jj}$ & 1.42 \\\hline
      $\eta({\rm untagged~jet})\big| p_T(jet)>9~$GeV & 0.78\\\hline
    \end{tabular}
  \end{center}
  \caption{The $\chi^2/ndf$ for all distributions in 
    $D^{*\pm}$-measurements shown.}
  \label{tab:dstarchi}
\end{table}
%%%%%%%%%%%%%%%%%%%%%%%%%%%%%%%%%%%%%%%%%%%%%%%%%%%%%%%%%%%%%%%%%%%%%%%%%%

\subsubsection{Inclusive $D^{*\pm}$ production}
%%%%%%%%%%%%%%%%%%%%%%%%%%%%%%%%%%%%%%%%%%%%%%%%%%%%%%%%%%%%%%%%%%%%%%%%%%
\begin{figure}
  \begin{center}
    \includegraphics[width=0.4\columnwidth]{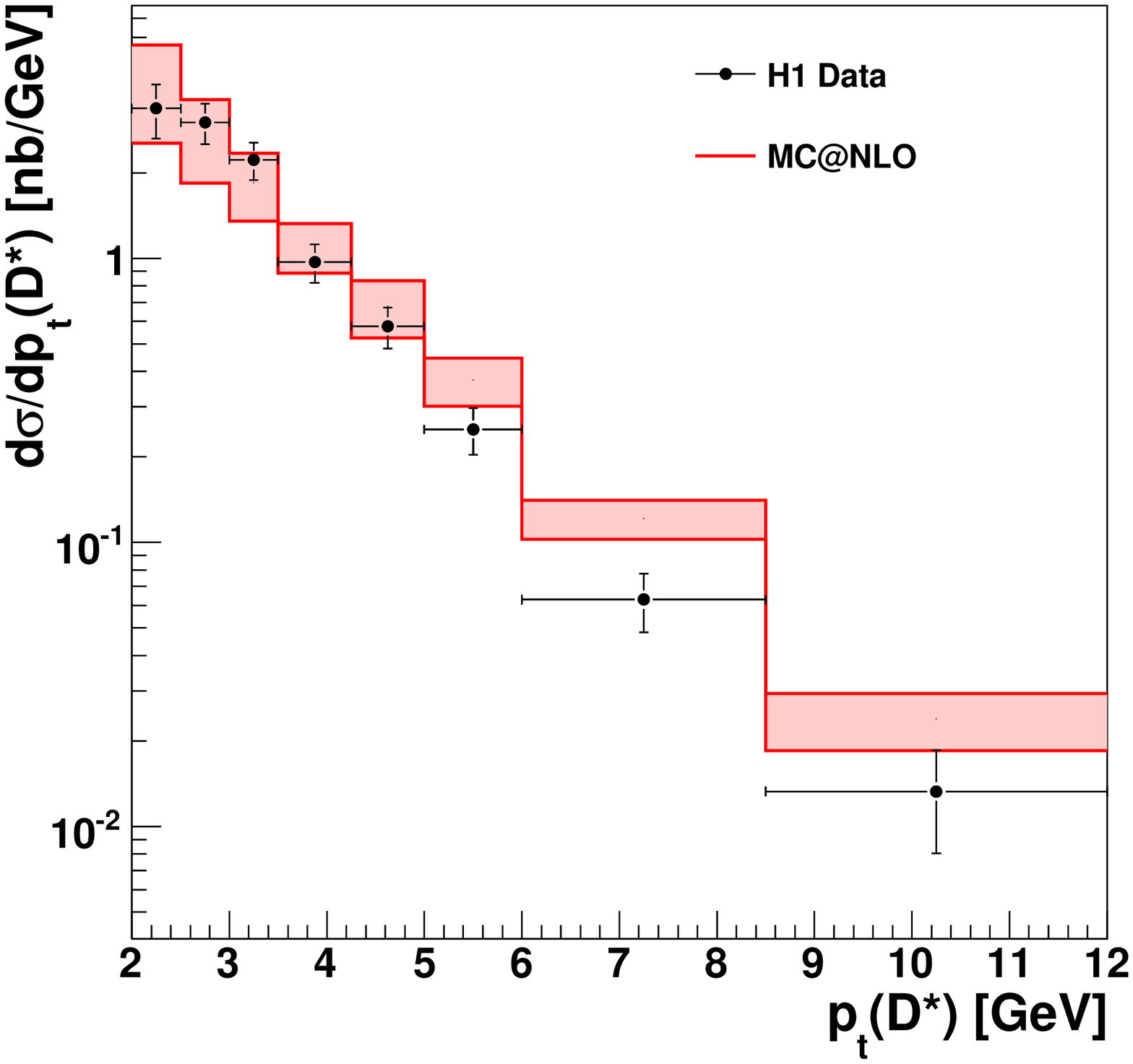}
    \includegraphics[width=0.4\columnwidth]{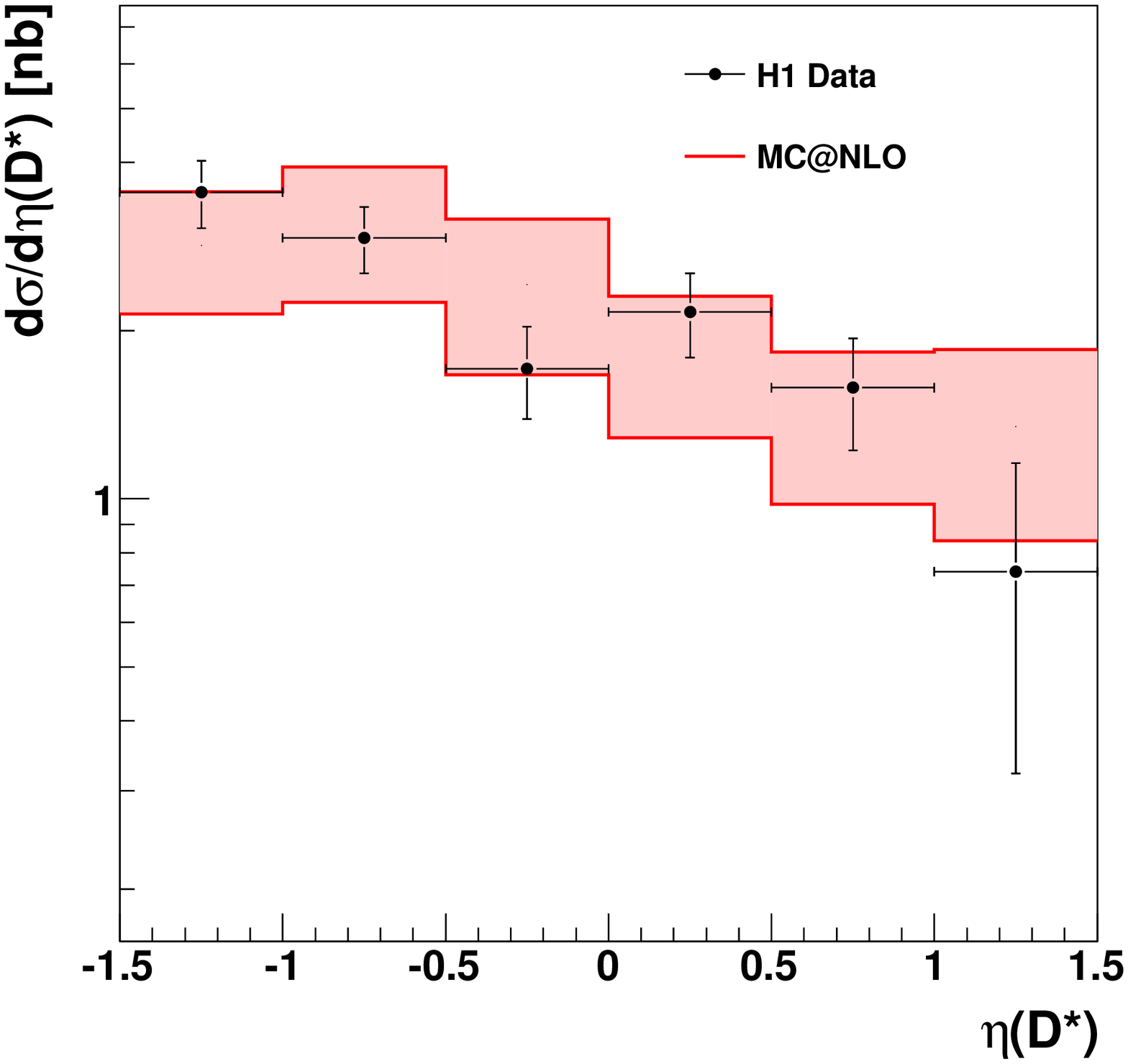}

    \includegraphics[width=0.4\columnwidth]{void.eps}
    \includegraphics[width=0.4\columnwidth]{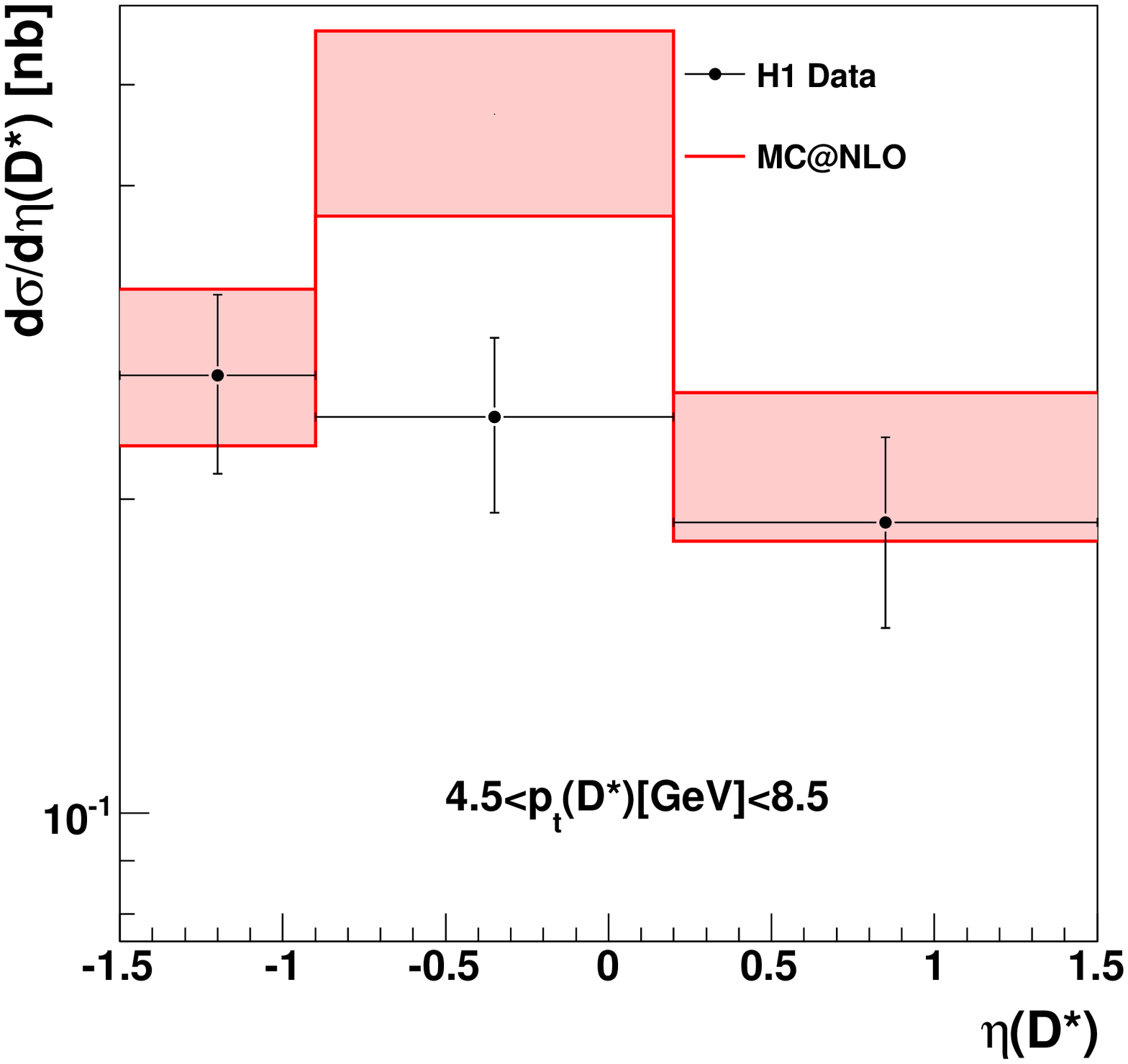}
  \end{center}
  \caption{Distributions of $p_T(D^*)$ and $\eta(D^*)$ of inclusive $D^*$
    measurements H1-06. The MC@NLO band includes full independent scale
    variations.}  
  \label{fig:incl.dpt}
\end{figure}
%%%%%%%%%%%%%%%%%%%%%%%%%%%%%%%%%%%%%%%%%%%%%%%%%%%%%%%%%%%%%%%%%%%%%%%%%%
In fig.~\ref{fig:incl.dpt} the $p_T(D^*)$ and $\eta(D^*)$ spectra are shown
for inclusive $D^*$ measurement. The H1-06 measurement is well described but  
the scale uncertainties in the MC@NLO prediction are rather
large in comparison with the experimental uncertainties. In the $p_T(D^*)$
distributions, these scale uncertainties are reduced for larger $p_T(D^*)$,
since the overall hardness of the process is increased.
In fig.~\ref{fig:incl.dpt}
the $\eta(D^*)$ distribution is also shown separately for $p_T(D^*)>4.5$~GeV,
where the scales uncertainties in MC@NLO are at the same level
as the experimental uncertainties. Here, two of the bins are well described,
but the overall shape is not. However, given the very large uncertainties
involved, the $\chi^2/ndf$ for this distribution is a mere 1.67.

\subsubsection{$D^{*\pm}$ plus jet production}
\begin{figure}
  \begin{center}
    \includegraphics[width=0.4\columnwidth]{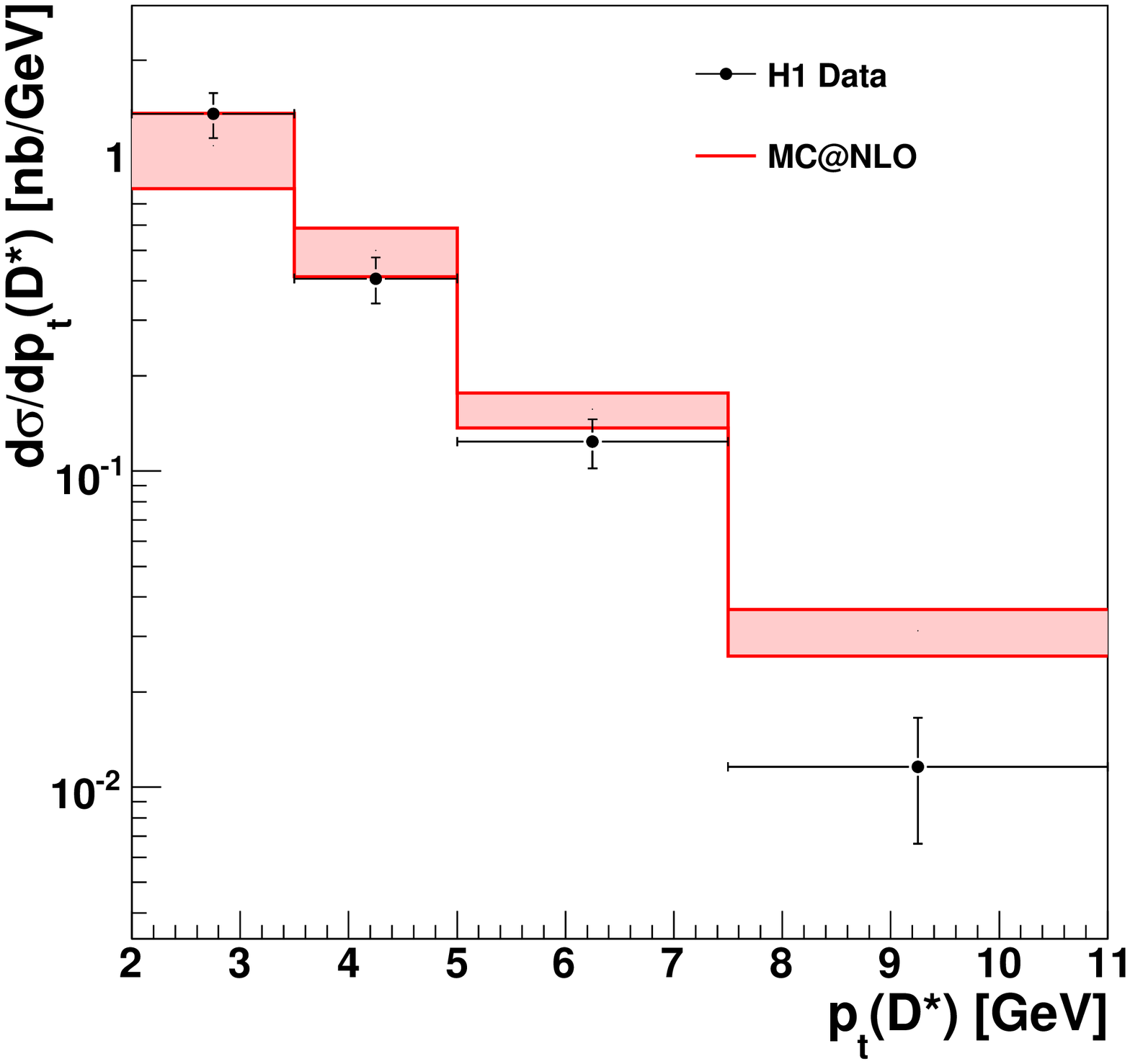}
    \includegraphics[width=0.4\columnwidth]{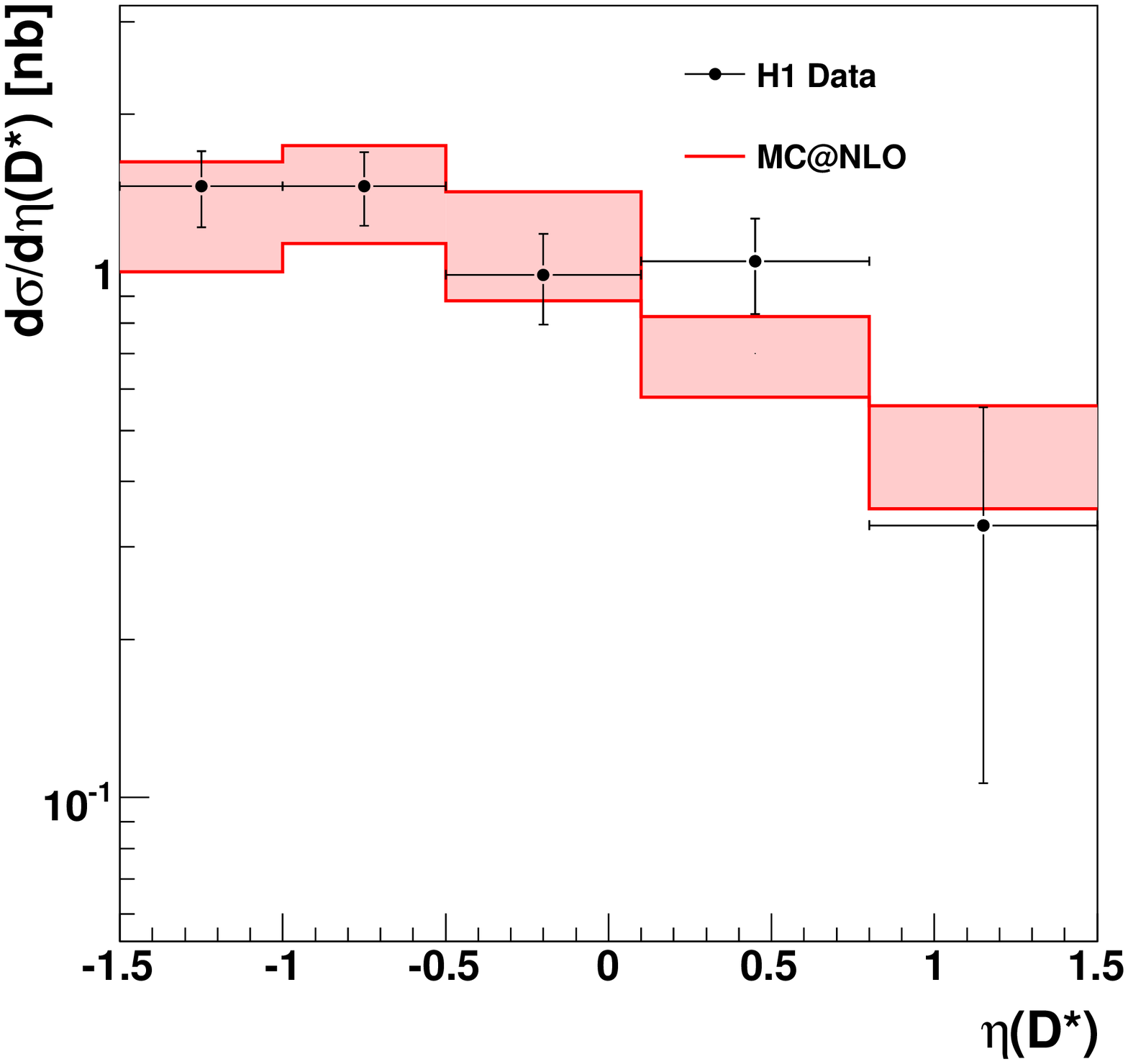}

    \includegraphics[width=0.4\columnwidth]{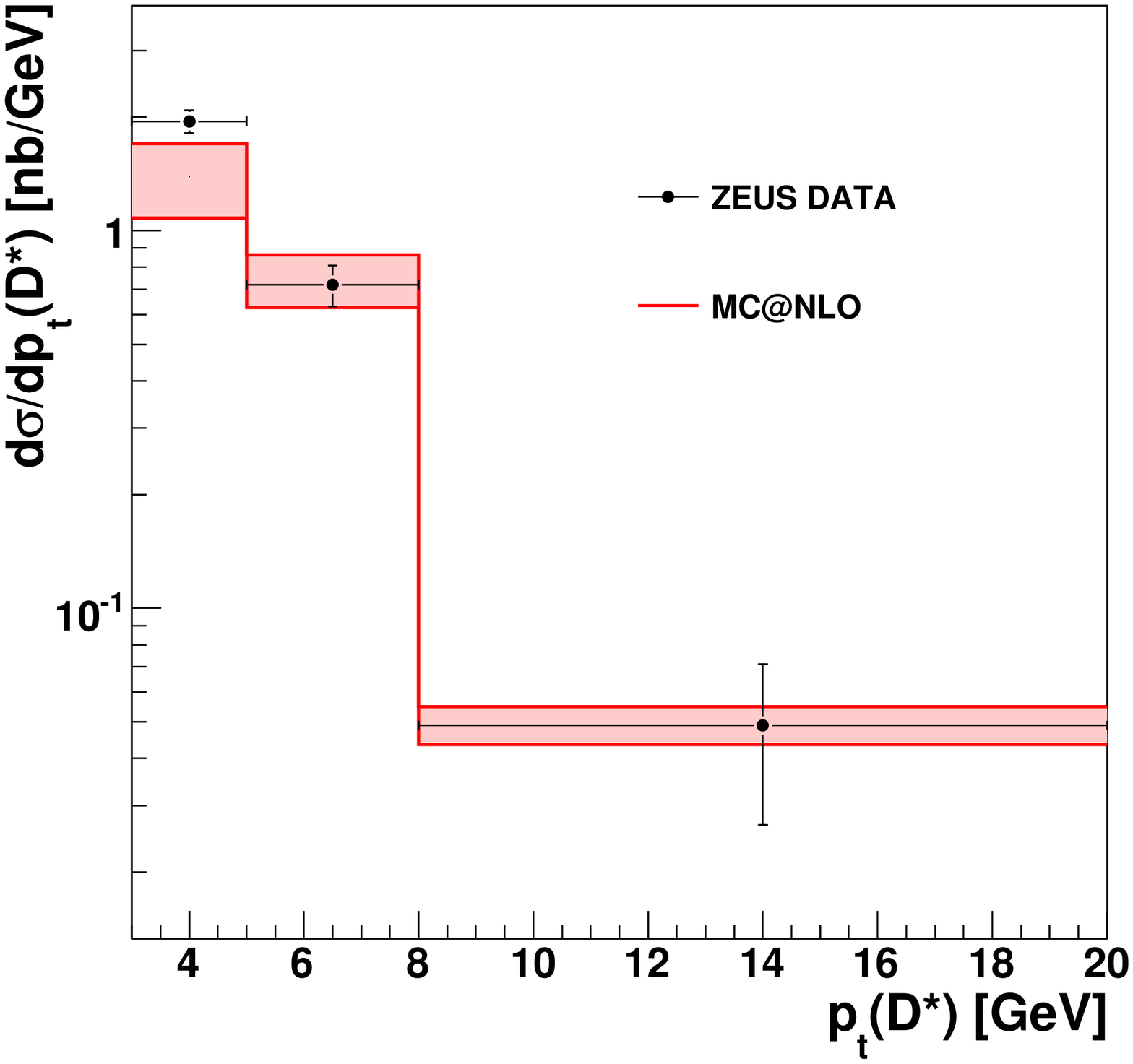}
    \includegraphics[width=0.4\columnwidth]{void.eps}
  \end{center}
  \caption{Distributions of $p_T(D^*)$ and $\eta(D^*)$ in the $D^*$ + jets 
    measurements from H1-06 as well as $p_T(D^*)$ from ZEUS-05.
    The MC@NLO band includes full independent scale variations.} 
  \label{fig:h1-06dstarjet}
\end{figure}

When demanding the tagging of a hard jet, as well as that of a $D^{*\pm}$ 
meson, the scale dependencies in MC@NLO are expected to be smaller
than for inclusive variables, owing to the extra hardness of the jet.
In fig.~\ref{fig:h1-06dstarjet}, the $p_T(D^*)$ and $\eta(D^*)$ distributions
from the H1-06 measurements are shown, when events with a hard jet with 
$p_T({\rm jet})>4$~GeV are chosen. The scale dependence
is indeed reduced. The data are still well described, even though the 
$p_T(D^*)$ spectrum is a bit harder in MC@NLO than in the H1-06 data,
while in ZEUS-05~\footnote{This $p_T(D^*)$ distribution is not presented in 
ref.~\cite{Chekanov:2005zg}. We have compiled it from other 
distributions binned in $p_T(D^*)$.} the opposite is observed.
%%%%%%%%%%%%%%%%%%%%%%%%%%%%%%%%%%%%%%%%%%%%%%%%%%%%%%%%%%%%%%%%%%%%%%%%%%
%%%%%%%%%%%%%%%%%%%%%%%%%%%%%%%%%%%%%%%%%%%%%%%%%%%%%%%%%%%%%%%%%%%%%%%%%%

\begin{figure}
  \begin{center}
    \includegraphics[width=0.4\columnwidth]{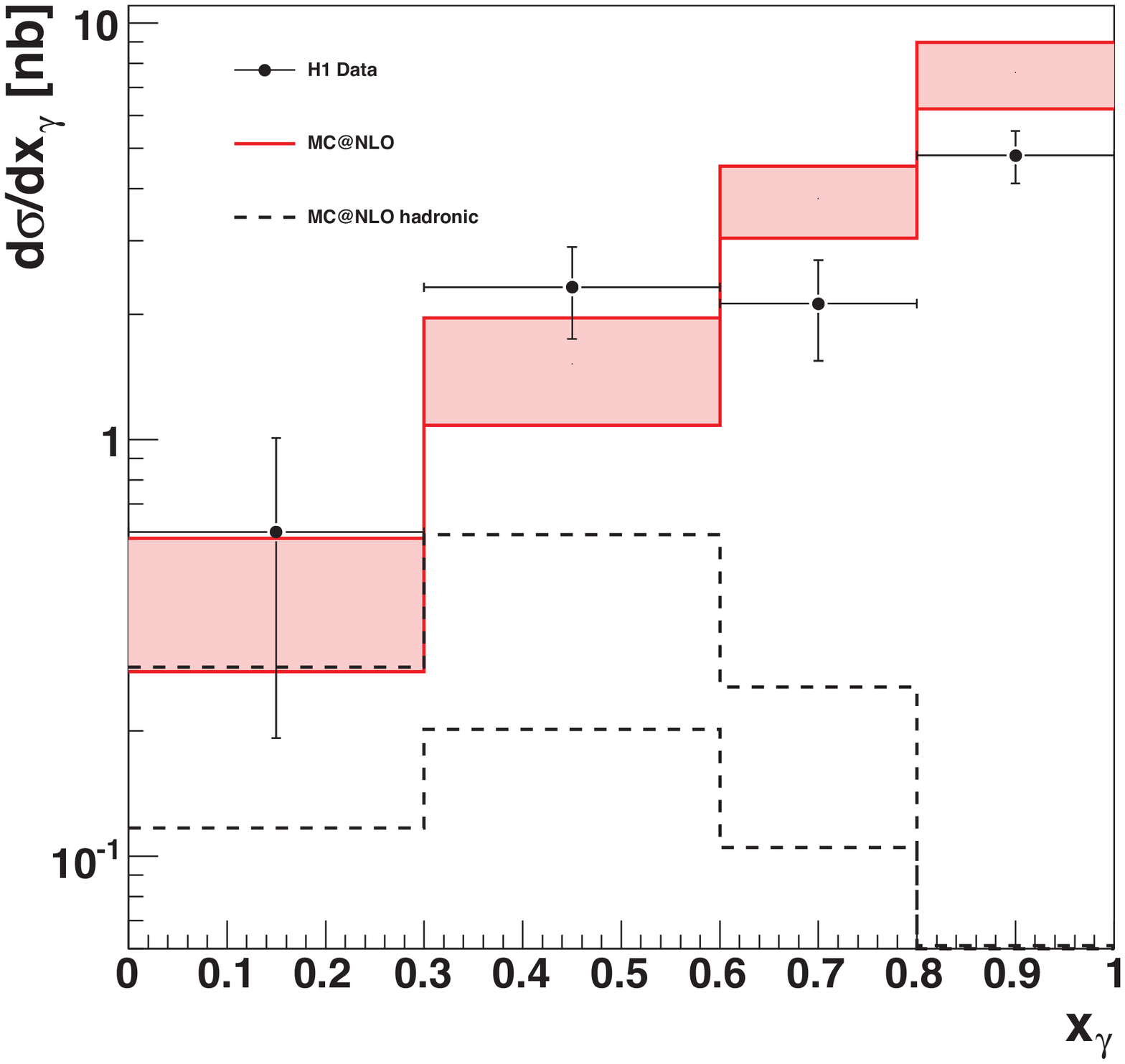}
    \includegraphics[width=0.4\columnwidth]{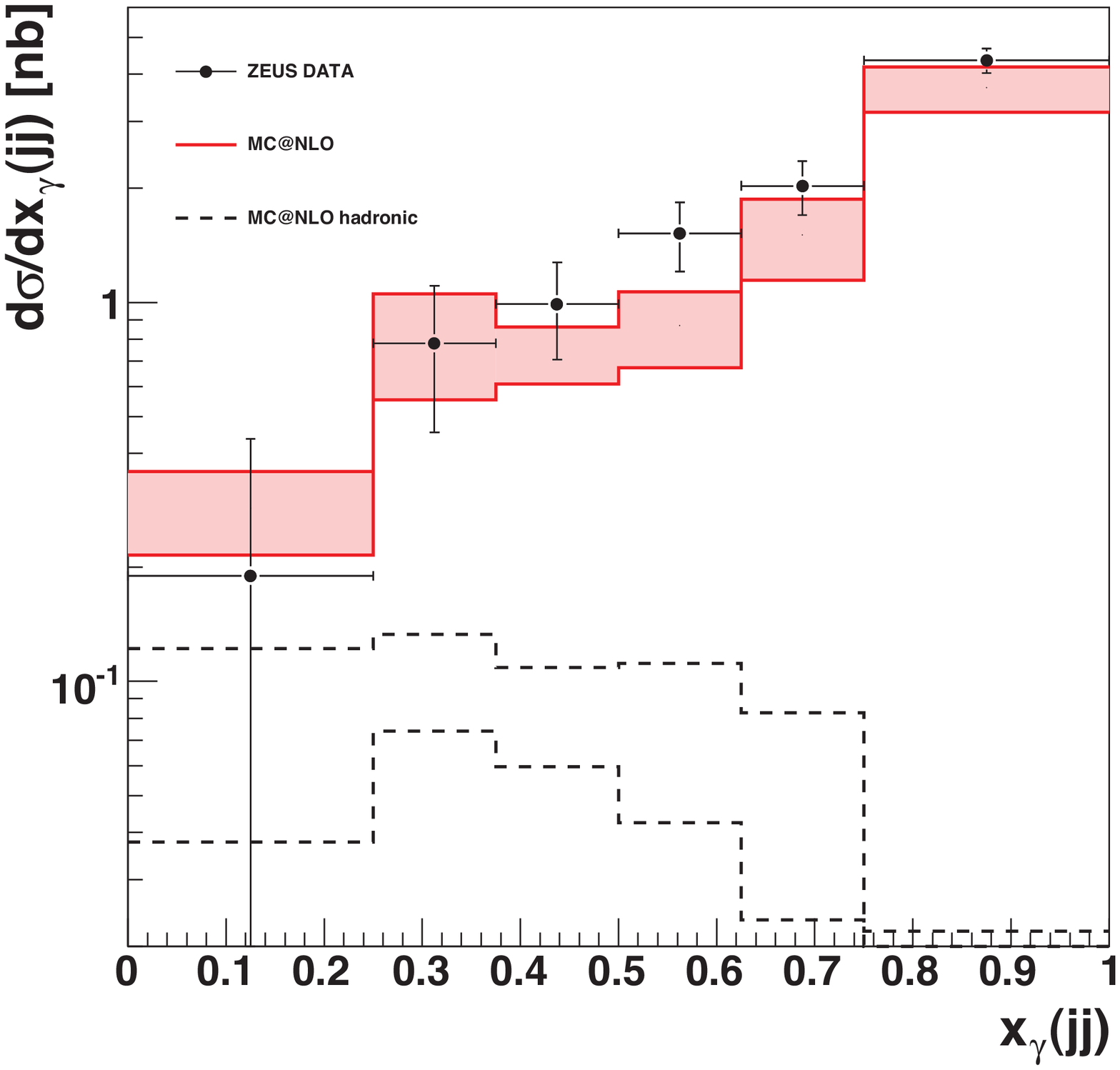}
  \end{center}
    \caption{Distributions of $x_\gamma^{\rm obs}({\rm jets})$ from H1-06 (left) and 
    ZEUS-05 (right). The MC@NLO band includes full independent scale variations.}
  \label{fig:xgamjj}
\end{figure}

Also, correlations between $D^*$ mesons and jets 
have been measured in H1-06 and ZEUS-05.
In fig.~\ref{fig:xgamjj}, distributions in $x_\gamma^{\rm obs}$ are shown. 
One can see that the contribution by the
hadronic part of MC@NLO is larger in the H1-06 measurement.
However, MC@NLO is one sigma
above the data for large $x_\gamma^{\rm obs}$ in this measurement, while
the whole spectrum is well described in the ZEUS-05 measurement. 
%%%%%%%%%%%%%%%%%%%%%%%%%%%%%%%%%%%%%%%%%%%%%%%%%%%%%%%%%%%%%%%%%%%%%%%%%%
%%%%%%%%%%%%%%%%%%%%%%%%%%%%%%%%%%%%%%%%%%%%%%%%%%%%%%%%%%%%%%%%%%%%%%%%%%

\begin{figure}
  \begin{center}
    \includegraphics[width=0.4\columnwidth]{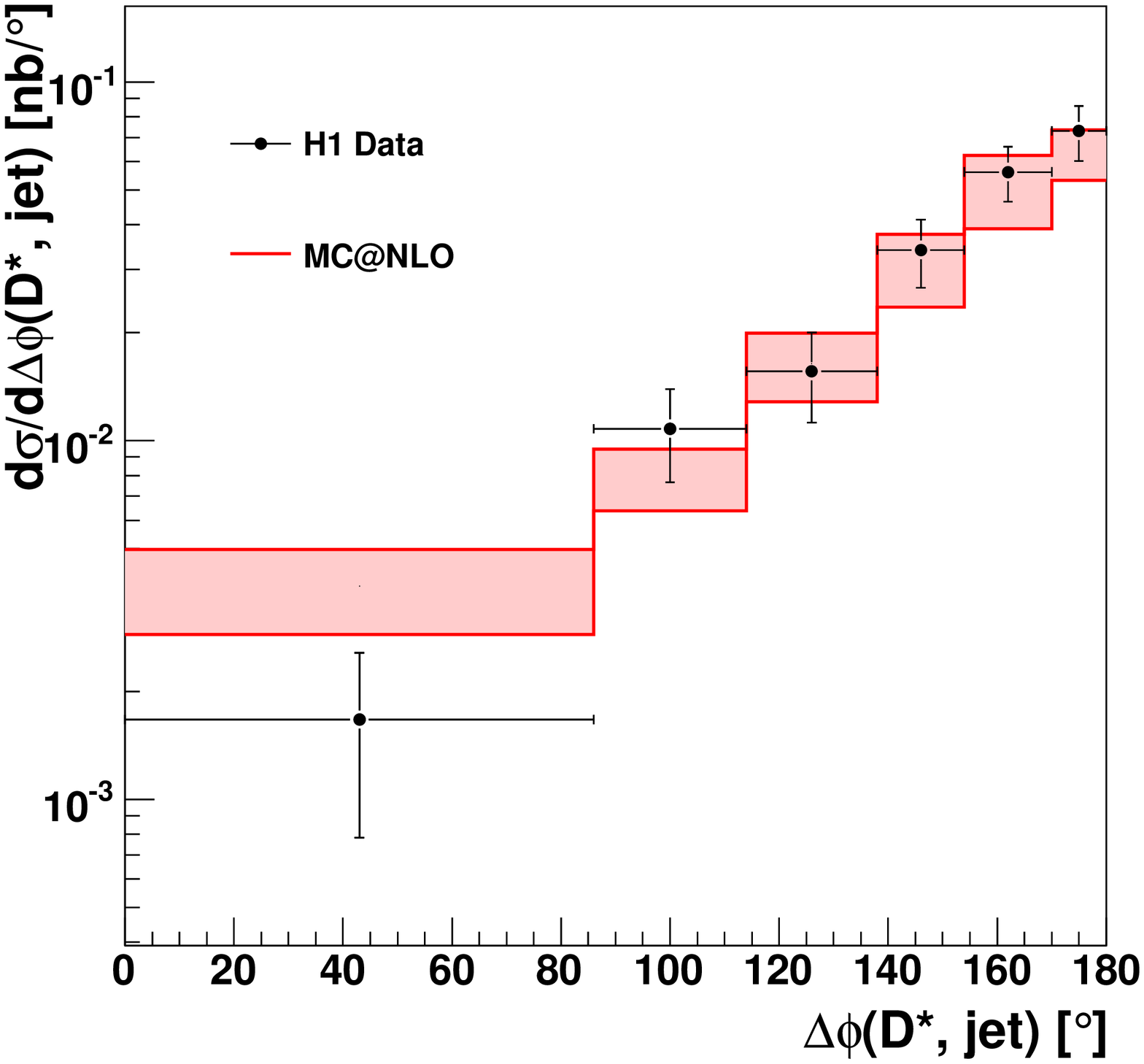}
    \includegraphics[width=0.4\columnwidth]{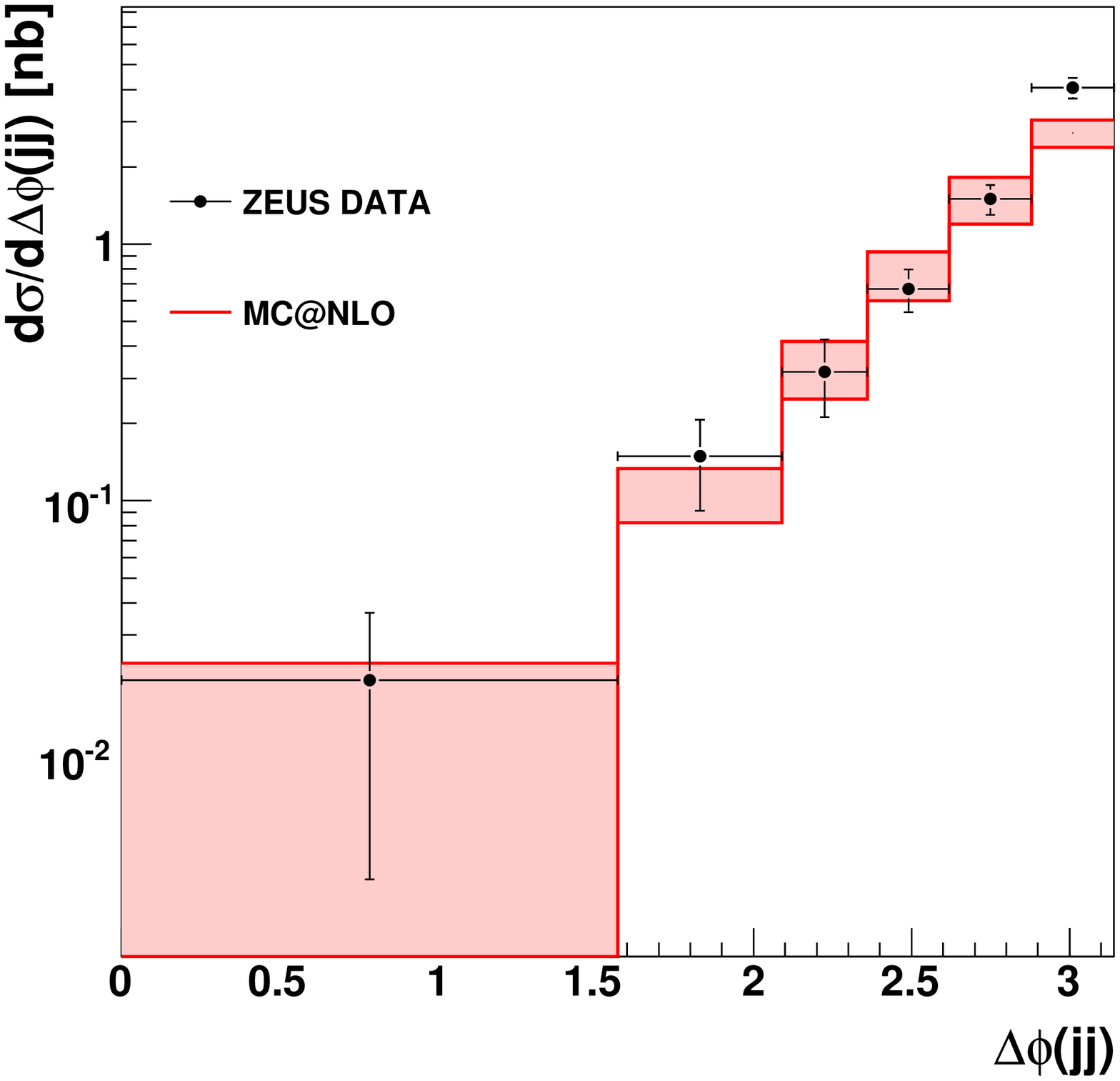}
  \end{center}
  \caption{Distributions of $\Delta\phi$ from H1-06 (left) and 
    ZEUS-05 (right). The MC@NLO band includes full independent scale variations.}
  \label{fig:ddeltaphi}
\end{figure}

In fig.~\ref{fig:ddeltaphi} the difference in azimuthal angle between the 
$D^*$ meson and the hardest jet not containing the $D^*$ (for H1-06), and
between the two hardest jets in a $D^*$ event (for ZEUS-05) are shown. 
One can see that
MC@NLO describes the data over the whole $\Delta\phi$ spectrum for both
analyses, something fixed order NLO calculations cannot do
(see e.g.~fig.~9 in ref.~\cite{Aktas:2006ry}, and fig.~11 in 
ref.~\cite{Chekanov:2005zg}).
%%%%%%%%%%%%%%%%%%%%%%%%%%%%%%%%%%%%%%%%%%%%%%%%%%%%%%%%%%%%%%%%%%%%%%%%%%
%%%%%%%%%%%%%%%%%%%%%%%%%%%%%%%%%%%%%%%%%%%%%%%%%%%%%%%%%%%%%%%%%%%%%%%%%%

\begin{figure}
  \begin{center}
    \includegraphics[width=0.4\columnwidth]{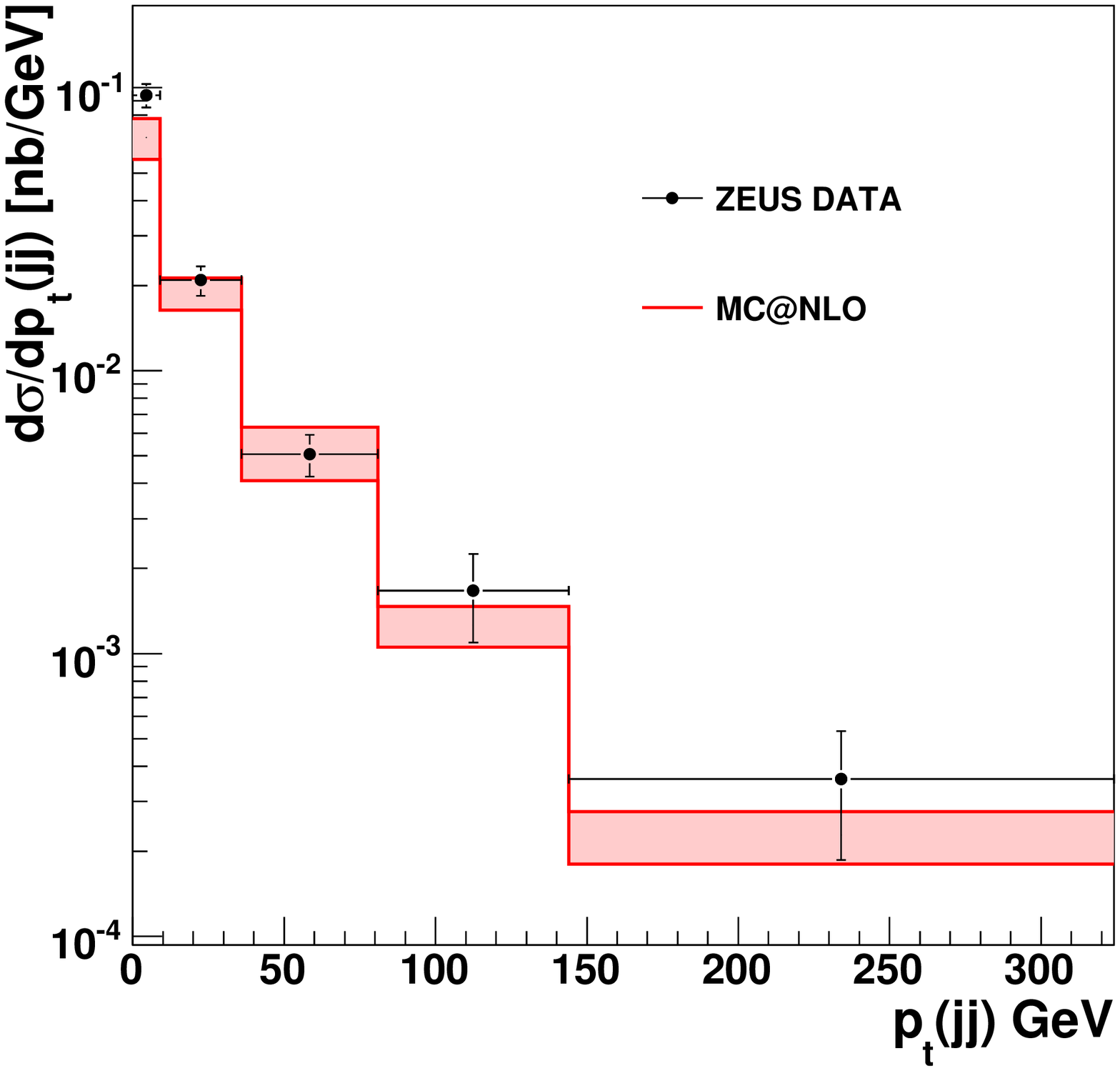}
    \includegraphics[width=0.4\columnwidth]{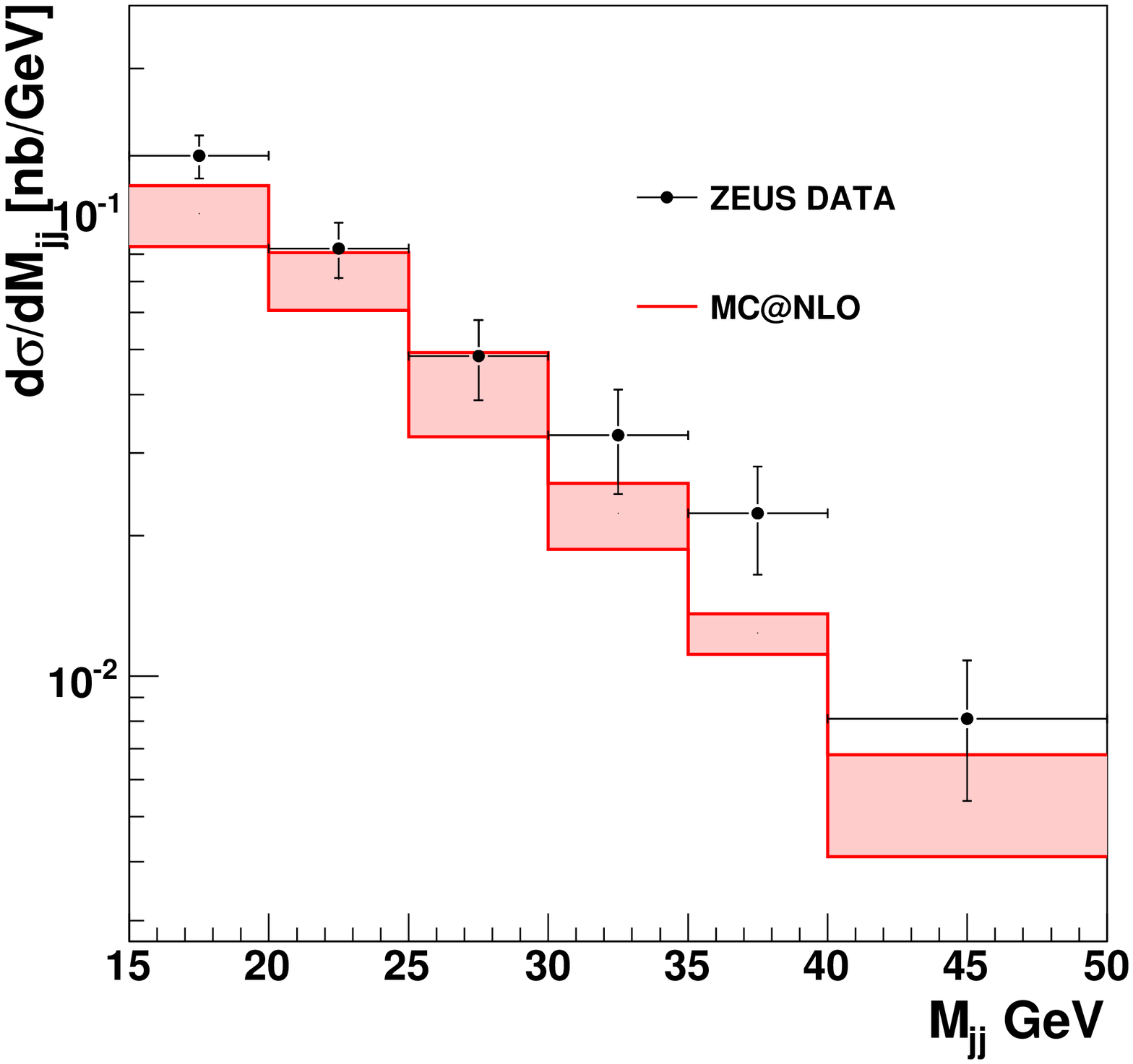}

    \includegraphics[width=0.4\columnwidth]{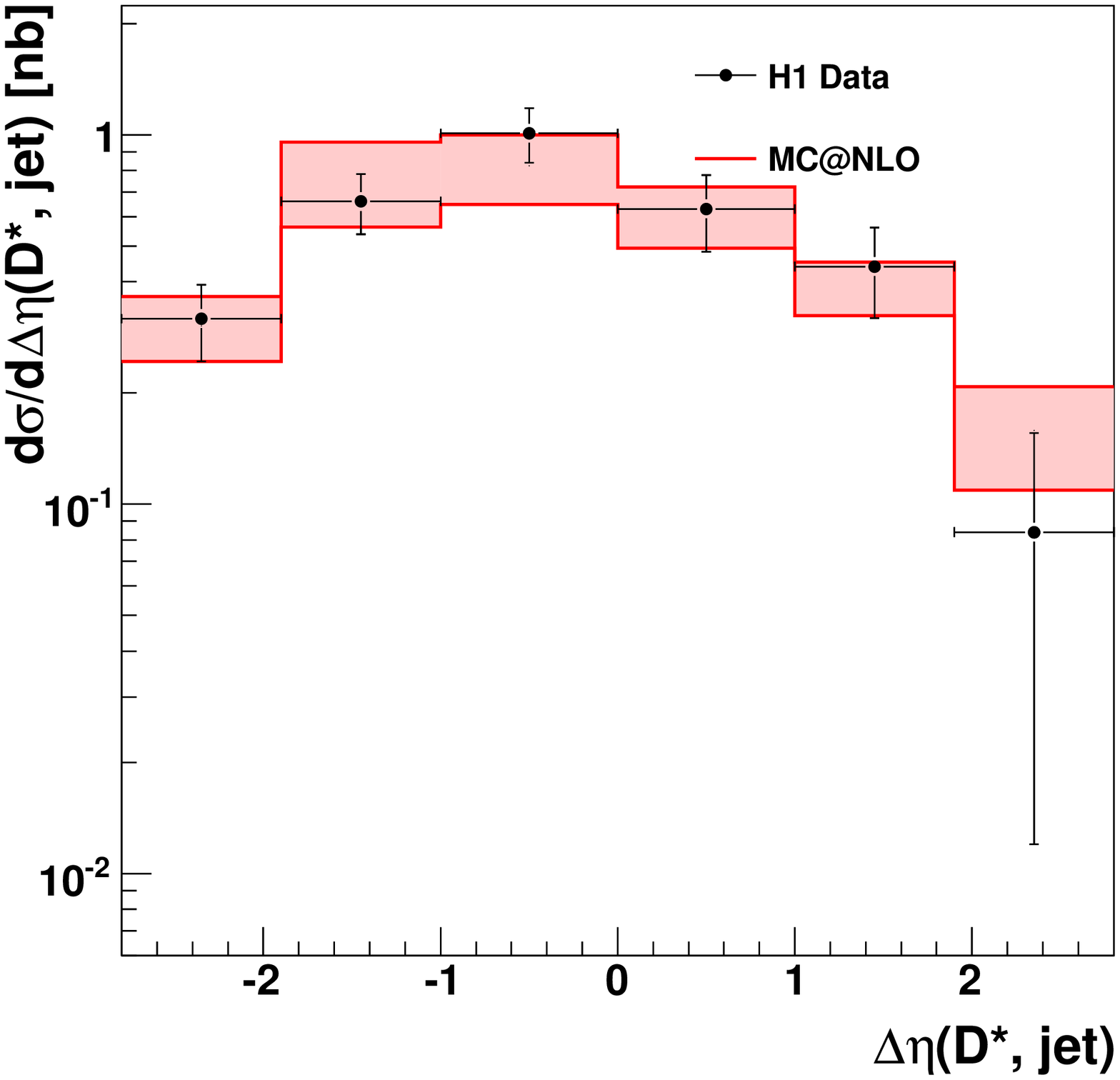}
    \includegraphics[width=0.4\columnwidth]{void.eps}
  \end{center}
  \caption{Distributions of $p_T({\rm jj})$ and $M_{\rm jj}$ from ZEUS-05
    and of $\Delta\eta(D^*, {\rm jet})$ from H1-06.
    The MC@NLO band includes full independent scale variations.}
  \label{fig:ptjjmjj}
\end{figure}

Another observable which is sensitive to higher order effects is the
$p_T$ of the two leading jets, which was measured in ZEUS-05, and shown
here in fig.~\ref{fig:ptjjmjj}. 
In this figure,  the distribution of the invariant mass 
of the two jets is also presented.
MC@NLO is seen to describe both these observables in a reasonable manner. 
%%%%%%%%%%%%%%%%%%%%%%%%%%%%%%%%%%%%%%%%%%%%%%%%%%%%%%%%%%%%%%%%%%%%%%%%%%
%%%%%%%%%%%%%%%%%%%%%%%%%%%%%%%%%%%%%%%%%%%%%%%%%%%%%%%%%%%%%%%%%%%%%%%%%%
Also, the difference in pseudo-rapidity between the $D^*$ and the hardest
other jet is sensitive to higher order radiations. This was measured in H1-06
as shown in the bottom plot of fig.~\ref{fig:ptjjmjj}. MC@NLO is describing
the spectrum in $\Delta\eta(D^*, {\rm jet})$ very well, with
$\chi^2/ndf=0.28$.

\section{Conclusions\label{sec:concl}}
In this letter, we have applied the MC@NLO formalism to the pointlike
photoproduction of heavy-quark pairs. Together with the already-available
$Q\bar{Q}$ hadroproduction code, this has been employed to carry out
a comparison between theoretical predictions and the data collected
by the H1 and ZEUS experiments at HERA, relevant to $b$- and $c$-hadron
observables. 

In particular, MC@NLO has been compared to five sets of measurements, 
three for $b$-flavoured hadrons, and two for $c$-flavoured hadrons.
All data have been shown to be described within one standard deviation 
by MC@NLO. It should be pointed out that MC@NLO predictions are absolute,
except for an overall rescaling needed
to obtain a branching ratio into muons (in the case of bottom production),
and into $D^*$'s (in the case of charm production) 
compatible with the world averages.

Although the overall agreement between theory and data appears to be
satisfactory, it should be kept in mind that the uncertainties involved
are sometimes quite large. For bottom production, theoretical errors
are at the same level or smaller than experimental
ones. On the other hand, for charm production the largest uncertainties
are those of MC@NLO, and therefore the comparisons carried out here
only loosely constrain perturbative QCD predictions. The situation
improves when hard jets are also part of the observable definitions.

\section{Acknowledgements}
The authors would like to thank Bryan Webber and Hannes Jung for great help
and useful discussions. T.T. was supported by DESY and by Hamburg University.
S.F. is on leave of absence from INFN, Sezione di Genova, Italy.

\end{document}